\documentclass[preprint,11pt]{elsarticle}

\usepackage{amssymb}
\usepackage{amsmath}
\usepackage{booktabs}
\usepackage{tipa}
\usepackage{xspace}
\usepackage{etoolbox}
\usepackage{microtype}
\usepackage{enumitem}
\usepackage{graphicx}
\usepackage{xurl}
\usepackage[margin=2.4cm]{geometry}
\usepackage[font=small,labelfont=bf,skip=4pt]{caption}

\makeatletter
\renewcommand\paragraph{\@startsection{paragraph}{4}{\z@}{0.8ex plus 0.3ex minus 0.2ex}{-0.5em}{\normalfont\normalsize\bfseries}}
\renewcommand\subsection{\@startsection{subsection}{2}{\z@}{1.6ex plus 0.4ex minus 0.2ex}{0.8ex plus 0.1ex}{\normalfont\large\bfseries}}
\renewcommand\subsubsection{\@startsection{subsubsection}{3}{\z@}{1.2ex plus 0.3ex minus 0.2ex}{0.6ex plus 0.1ex}{\normalfont\normalsize\bfseries}}
\let\cslparagraph\paragraph
\renewcommand\paragraph[1]{\cslparagraph{#1:}}
\makeatother

\setlist[itemize]{noitemsep,topsep=2pt}
\AtBeginEnvironment{table}{\small}
\AtBeginEnvironment{table*}{\small}

\journal{Computer Speech \& Language}

\newcommand{\auc}{AUC\xspace}
\newcommand{\ap}{AP\xspace}
\newcommand{\eer}{EER\xspace}
\newcommand{\balacc}{BalAcc\xspace}

\begin{document}

\begin{frontmatter}

\title{An evaluation study of consonants in synthetic speech: fricatives stand out}

\author[1]{Sriyugesh Bhyravajjula\corref{cor1}}
\ead{bsriyugesh@gmail.com}
\author[2]{Ayushi Pandey}
\author[3,4]{Kevin Tang}
\cortext[cor1]{Corresponding author}

\affiliation[1]{organization={Mahindra University},
               country={India}}
\affiliation[2]{organization={Karya},
               country={India}}
\affiliation[3]{organization={Department of English Language and Linguistics, Institute of English and American Studies, Faculty of Arts and Humanities, Heinrich Heine University D\"usseldorf},
               country={Germany}}
\affiliation[4]{organization={Department of Linguistics, University of Florida},
               country={United States of America}}

\begin{abstract}
Text-to-speech (TTS) quality is conventionally assessed at the utterance level, which does not localise where synthetic speech departs from human speech.
Earlier work on one speaker and older systems implicated fricatives, and we test whether this generalises.
We compare human speech with 11 TTS systems, three of which clone a voice, on two corpora (VCTK with 30 speakers, and LJSpeech) and five phoneme classes: fricatives, stops, affricates, nasals and approximants.
Paired statistics and one classifier per speaker and system quantify the differences.
Six acoustic features (five spectral or energy features and duration) give a mean area under the ROC curve (AUC) of 0.95 for fricatives, against 0.85 for affricates, 0.83 for stops, 0.76 for approximants and 0.71 for nasals.
With these features, fricatives lead in every context and voicing category and in 15 of 16 system configurations.
The lead persists for the voice-cloning systems (0.95) and without root-mean-square (RMS) amplitude (0.94).
Class-neutral features raise the separability of the other classes: fricatives stay first under linear-frequency cepstral coefficients (LFCC; 0.98, against 0.92 to 0.96) and tie with nasals under mel-frequency cepstral coefficients (MFCC; 0.91 and 0.88).
Fricative cues also generalise to unseen speakers better than those of other classes (0.86, against 0.55 to 0.77).
Cues do not transfer between corpora: obstruent classifiers score below chance on the other corpus, and sonorant classifiers at chance.
Fricatives are therefore the most discriminative consonant class under the class-specific and LFCC representations, but the cues must be learned in the target domain.
\end{abstract}


\end{frontmatter}

\section{Introduction}
\label{sec:introduction}

Modern text-to-speech (TTS) systems produce speech of high perceptual quality.
Evaluation nevertheless operates mainly at the utterance level, for example through listener ratings.
An utterance-level score does not indicate where in the signal a synthetic voice still departs from a human one.

Speech comprises segments, and segments of different classes arise from different articulatory mechanisms.
Nasals and approximants are voiced and resonant.
Obstruents are produced by blocking or constricting the airflow.
Fricatives are produced by turbulent noise generated at a narrow constriction.
A TTS system must model all of these mechanisms.
It may model some less faithfully than others.
If so, one class may be the most reliable indicator of synthesis.

Earlier work on obstruents in TTS points in this direction \cite{pandey2022production,pandey2023natural,bhyravajulla2025fricatives}.
It compared acoustic measurements of synthetic and human obstruents, and of fricatives in particular.
Our earlier study of fricatives was limited to one speaker and four systems \cite{bhyravajulla2025fricatives}.
A difference observed for a single voice need not hold for other voices.
Four systems also exclude recent architectures, such as those that clone a voice from a short recording.

This paper tests whether fricatives are indeed the most separable class.
We examine 31 speakers from two corpora and 11 TTS systems, three of which clone a voice from a reference recording.
We compare five phoneme classes: fricatives, stops, affricates, nasals and approximants.

We address this question in three experiments.
Experiment 1 quantifies how obstruents in synthetic speech differ from those in human speech.
Experiment 2 performs the same analysis for sonorants.
Experiment 3 trains classifiers on both groups.
Classifier performance indicates how well the features of a class separate synthetic from human tokens.

The paper makes three contributions.
\begin{itemize}
  \item It shows that fricatives are the most separable class under six class-specific features (mean area under the ROC curve, \auc, 0.95, against 0.71 to 0.85) and under linear-frequency cepstral coefficients (LFCC; 0.98, against 0.92 to 0.96); under mel-frequency cepstral coefficients (MFCC) they tie with nasals (0.91 and 0.88).
  \item It shows that this lead persists for three voice-cloning systems and when the level-dependent RMS feature is removed.
  \item Under a single design that covers five phoneme classes, 31 speakers and 11 systems, it shows that cues do not transfer between corpora, so they must be learned for each target domain.
\end{itemize}

Section~\ref{sec:related} reviews related work.
Section~\ref{sec:methodology} describes the data and the methods.
Section~\ref{sec:results} reports the results.
Section~\ref{sec:discussion} discusses what the fricative result means and where it stops.
Section~\ref{sec:conclusion} concludes.

\section{Related work}
\label{sec:related}

Segmental studies compare acoustic measurements of synthetic and human consonants.
Pandey et al.\ studied obstruents in WaveNet and older TTS systems \cite{pandey2022production}.
They also studied place classification of human and Tacotron fricatives \cite{pandey2023natural}.
Our earlier study repeated that evaluation on a second speaker, the LJSpeech speaker, and found that voiceless fricatives of VITS deviate from human speech and that flow-based systems deviate less \cite{bhyravajulla2025fricatives}.

Work on synthetic speech detection has also used phoneme-level information.
Zhang et al.\ detect deepfakes from inconsistencies of phoneme-level features across a sequence \cite{zhang2025phoneme}.
PhonemeDF provides parallel human and synthetic speech segmented by phoneme \cite{nallaguntla2026phonemedf}.
It relates the divergence of phoneme distributions to classifier performance.
It reports that consonants are generally more discriminative than vowels.
Among vowels, diphthongs are the most discriminative.
Among consonants, fricatives and plosives lead.
Chhibber et al.\ rank phonetic categories inside a phoneme-guided detector \cite{chhibber2026phonetically}.
They report that stops, fricatives, affricates, nasals and silence-boundary closures are the most discriminative categories, and that periodic vowels and semivowels rank lower.
Temmar et al.\ analyse human and synthetic speech at the phoneme and word level with HuBERT embeddings \cite{temmar2025phonetic}.
PhonemeDF summarises their analysis as showing that diphthongs and fricatives deviate most from human speech \cite{nallaguntla2026phonemedf}.
Sivaraman et al.\ train an AASIST detector separately on voiced and unvoiced regions of speech \cite{sivaraman2025voiced}.
Unvoiced regions separate synthetic from bona fide speech better (equal error rate 6.62\%), and fusing both regions lowers the error rate to 5.82\%.
Unvoiced regions contain the fricatives, so this result is compatible with ours.

These studies and ours differ in what they compare.
They use detectors, learned representations or distances between distributions.
We compare five consonant classes with fixed acoustic features, inside each speaker and system.
Our design uses two corpora and 11 systems, three of which clone a voice from a reference recording.

The rankings in the literature depend on the representation.
PhonemeDF finds that handcrafted spectral features highlight fricatives and diphthongs, while self-supervised embeddings emphasise plosives \cite{nallaguntla2026phonemedf}.
Our result concerns handcrafted spectral features.
We return to this point in Section~\ref{sec:discussion}.

\section{Methodology}
\label{sec:methodology}

\subsection{Data and experimental setup}
\label{subsec:common}

\subsubsection{Human speech}

We use human speech from two corpora, VCTK and LJSpeech.
VCTK \cite{veaux2017cstr} is a multi-speaker English corpus.
We use 30 of its speakers, 15 female and 15 male.
A multi-speaker corpus lets us test whether a difference holds across voices, and not only for one voice.

LJSpeech \cite{ljspeech17} is a single-speaker corpus.
Earlier work on synthetic fricatives used it \cite{bhyravajulla2025fricatives}.
That work compared four TTS systems on LJSpeech with Dunn's test and Cliff's delta.
We include LJSpeech so that we can extend that work with more systems and a paired analysis.

Together the two corpora supply 117{,}929 human fricative tokens, 96{,}282 stop tokens, 4{,}947 affricate tokens and 107{,}253 sonorant tokens.
All recordings are resampled to 16\,kHz.
For VCTK we use the first microphone recording of each utterance.

\subsubsection{Synthetic speech}

We use pretrained TTS models. Eight are trained on VCTK or LJSpeech, and three clone a voice from a reference recording.
Table~\ref{tab:systems} lists the systems.

\begin{table}[!htbp]
\centering
\caption{Synthetic systems. The VCTK column gives the number of speakers synthesised, and the LJSpeech column marks the systems used with the single LJSpeech speaker. ``HiFi-GAN'' means that a mel-spectrogram model is paired with a HiFi-GAN vocoder.}
\label{tab:systems}
\begin{tabular}{llccc}
\toprule
System & Type & Waveform generation & VCTK & LJSpeech \\
\midrule
DiffGAN-TTS     & Diffusion GAN        & HiFi-GAN            & 30 & yes \\
FastSpeech~2    & Non-autoregressive   & HiFi-GAN            & 30 & yes \\
Tacotron~2      & Autoregressive       & HiFi-GAN            & 30 & yes \\
Transformer~TTS & Autoregressive       & HiFi-GAN            & 30 & yes \\
VITS            & End to end           & Part of the system  & 30 & yes \\
GlowTTS         & Flow-based           & HiFi-GAN            & -- & yes \\
GradTTS         & Diffusion            & HiFi-GAN            & -- & yes \\
PFlow           & Flow-based           & HiFi-GAN            & -- & yes \\
\midrule
CosyVoice~2     & Zero-shot voice cloning & Part of the system & 10 & -- \\
FishSpeech      & Zero-shot voice cloning & Part of the system & 10 & -- \\
Qwen3-TTS       & Zero-shot voice cloning & Part of the system & 10 & -- \\
\bottomrule
\end{tabular}
\end{table}

Five systems are trained on VCTK and synthesise all 30 selected speakers: DiffGAN-TTS \cite{liu2022diffgantts}, FastSpeech~2 \cite{ren2021fastspeech2}, Tacotron~2 \cite{shen2018tacotron2}, Transformer~TTS \cite{li2019transformertts} and VITS \cite{kim2021vits}.
The speakers are chosen from the 108 VCTK speakers that all five systems support.
We rank the speakers by number of utterances within each gender.
The first group has the five female and five male speakers with the most utterances.
The second group has the next ten female and ten male speakers.

VITS generates the waveform itself from its public checkpoint.
The other four systems pair a mel-spectrogram model with a HiFi-GAN vocoder \cite{kong2020hifigan} trained on VCTK.
FastSpeech~2, Tacotron~2 and Transformer~TTS run through the ESPnet2 toolkit \cite{watanabe2018espnet,hayashi2021espnet2tts} with its published VCTK checkpoints.
They also receive a speaker x-vector and a style reference clip from the target speaker.
DiffGAN-TTS runs from its own repository.

Three recent systems clone a voice from a reference recording: CosyVoice~2 \cite{du2024cosyvoice2}, FishSpeech \cite{liao2024fishspeech} and Qwen3-TTS \cite{hu2026qwen3tts}.
They are not trained on the target speakers.
Each is conditioned on a reference recording of the target speaker and its transcript.
The longest recording of each speaker is the reference, and it is withheld from synthesis.
The checkpoints are CosyVoice2-0.5B, OpenAudio S1-mini (FishSpeech) and Qwen3-TTS-12Hz-1.7B-Base.
The reference recording stays among the human recordings, where it makes up a small share of the human tokens of these speakers.
It has no synthetic counterpart, so it does not enter the paired statistics.

Every synthesised utterance uses the transcript of a human recording by the same speaker.
Each synthetic token therefore has a human counterpart with the same text.

Eight systems are used with LJSpeech.
FastSpeech~2, Tacotron~2, Transformer~TTS, DiffGAN-TTS, GlowTTS \cite{kim2020glowtts}, GradTTS \cite{popov2021gradtts} and PFlow \cite{kim2023pflow} each use a HiFi-GAN vocoder.
VITS generates the waveform itself.
GlowTTS, GradTTS and PFlow are used only with LJSpeech.
The synthesised sentences are the held-out test sentences of each system, as specified in its repository \cite{bhyravajulla2025fricatives}.
The number of analysed utterances differs by system, from 99 for Transformer~TTS to 1{,}326 for FastSpeech~2 and Tacotron~2.

In total there are 16 synthetic configurations: eight on VCTK and eight on LJSpeech.
They use 11 distinct system names, because five systems (DiffGAN-TTS, FastSpeech~2, Tacotron~2, Transformer~TTS and VITS) run on both corpora and are counted once per corpus.

Two of the three voice-cloning systems, CosyVoice~2 and FishSpeech, needed adaptation to run with the released checkpoints, so our results describe these builds.
The first CosyVoice~2 implementation lacked a decoder class and a configuration parameter that the checkpoint requires, so we replaced it with the upstream repository.
The FishSpeech checkpoint loads only with the repository fixed to one commit, because the current release has a tokenizer that does not support the checkpoint format and the nearest tagged release has a model configuration without the fields the checkpoint needs.
CosyVoice~2 ran on a CPU, and FishSpeech and Qwen3-TTS ran on an Apple-silicon GPU.
These systems cover only the first group of ten speakers.

\subsubsection{Segmentation, phoneme classes and contexts}

Phoneme boundaries come from forced alignment with the Montreal Forced Aligner \cite{mcauliffe2017montreal}.
We align human and synthetic recordings in the same way.
We use version 3.2.3 with its pretrained US English ARPA acoustic model and pronunciation dictionary.

Phonemes are grouped into manner classes.
Fricatives are /f, v, s, z, \textipa{S}, \textipa{Z}, \textipa{T}, \textipa{D}, h/.
Stops are /p, b, t, d, k, g/.
Affricates are /\textipa{tS}, \textipa{dZ}/.
Nasals are /m, n, \textipa{N}/.
Approximants are /l, \textipa{r}, w, j/.
Fricatives, stops and affricates are the obstruents.
Nasals and approximants are the sonorants.
We group /h/ with the voiceless fricatives.

Each phoneme is assigned a context from its neighbours in the lexicon pronunciation of its word.
A phoneme followed by a vowel is labelled CV.
A phoneme preceded by a vowel is labelled VC.
A phoneme between two vowels is labelled VC.
We adopted this convention so that every phoneme with a preceding vowel falls in one context, and we did not test the alternative.
A phoneme with no vowel neighbour in the word is labelled isolated.

\subsubsection{Dataset composition}

This section describes the datasets that the experiments use.
Table~\ref{tab:data} gives the number of human and synthetic tokens in each analysed dataset, for the four phoneme groups.
Table~\ref{tab:mv-data} gives the tokens and phonemes for each manner class and voicing category.
Table~\ref{tab:data-system} gives the synthetic fricative tokens and utterances of each system.
Table~\ref{tab:context-share} gives the share of human tokens in each context and the share of voiced tokens.

\begin{table}[!htbp]
\centering
\caption{Tokens in the analysed tables, by corpus and phoneme group. Utterances are the human utterances that contain the phoneme group. The ratio is synthetic to human tokens. LJSpeech rows cover eight synthetic systems. The sonorant counts are before the formant validity filter. After the filter, 44{,}479 human and 210{,}191 synthetic nasal tokens remain, and 55{,}495 human and 274{,}565 synthetic approximant tokens remain.}
\label{tab:data}
\begin{tabular}{llrrrc}
\toprule
Corpus & Phoneme group & Human tokens & Synthetic tokens & Utterances & Ratio \\
\midrule
VCTK & Fricatives & 89{,}332 & 383{,}818 & 13{,}369 & 4.3 \\
 & Stops & 65{,}928 & 400{,}101 & 13{,}339 & 6.1 \\
 & Affricates & 3{,}417 & 20{,}945 & 2{,}831 & 6.1 \\
 & Sonorants & 73{,}969 & 449{,}539 & 13{,}503 & 6.1 \\
\midrule
LJSpeech & Fricatives & 28{,}597 & 64{,}133 & 2{,}305 & 2.2 \\
 & Stops & 30{,}354 & 68{,}645 & 2{,}304 & 2.3 \\
 & Affricates & 1{,}530 & 3{,}399 & 1{,}078 & 2.2 \\
 & Sonorants & 33{,}284 & 75{,}321 & 2{,}308 & 2.3 \\
\bottomrule
\end{tabular}
\end{table}

Synthetic tokens outnumber human tokens because one set of human recordings is the reference for every system.
LJSpeech has more human utterances (2{,}305 with fricatives) than any one system synthesised (at most 1{,}326).
Some human utterances therefore have no synthetic counterpart.
These utterances do not enter the paired statistics.

Voiced and voiceless fricatives are nearly balanced.
Stops are less balanced: voiceless stops make up 0.63 of the human stop tokens.
The LJSpeech speaker contributes far more human tokens than any single VCTK speaker (Table~\ref{tab:data}).
Affricates are rare.
There are only 3{,}417 human affricate tokens in VCTK and 1{,}530 in LJSpeech, and the isolated context is the smallest.

\subsubsection{Human reference}

The human reference for a synthetic recording is the human recording of the same speaker and sentence.
For LJSpeech it is the LJSpeech recording of the same test sentence.
One set of human recordings is therefore reused for every system synthesised for that speaker.

\subsubsection{Pairing and class balance}

The statistical tests of Experiments 1 and 2 compare paired tokens.
A synthetic token is paired with a human token when both have the same sentence, word position, word and phoneme position.
A synthetic token with no such human token is not paired and is left out of the test.
Table~\ref{tab:pairing} gives the share of synthetic tokens that have a human counterpart.
The share is at least 97\% in every row.

The classifiers of Experiment 3 do not use pairs.
A classifier cell holds the synthetic tokens of one system and all human tokens of the same speaker.
The two classes therefore differ in size.
We give the synthetic class a training weight equal to the ratio of human to synthetic training tokens.
The paired tests use only matched tokens, so the ratio of the two classes does not enter them.

\subsubsection{Analysis code}
The analysis code was written with the assistance of Claude (Anthropic) and reviewed by the authors.
All reported numbers are produced by this code.

\subsection{Experiment 1: statistical analysis of obstruents}
\label{subsec:p1-method}

\subsubsection{Setup}

Experiment 1 tests which acoustic properties separate synthetic from human obstruents, and how consistently.
The obstruents are fricatives, stops and affricates.
We analyse all three with one procedure, so that the manner classes can be compared directly.
The experiment covers 188 speaker--system pairs.
On VCTK there are 180 pairs: five conventional systems for 30 speakers and three cloning systems for 10 speakers.
On LJSpeech there are eight pairs.
We run 26{,}525 obstruent tests over the eight features below.
Each test is run separately for one context, manner class, voicing category and feature.

\subsubsection{Features}

Eight acoustic features describe each token.
First, the token is pre-emphasised with a coefficient of 0.97.
Then it is high-pass filtered at 200\,Hz with a fifth-order Butterworth filter, applied forwards and backwards.
Tokens too short for the filter's padding are not filtered.

The spectral features use the central 40\,ms of the token.
We chose this length for convenience and did not compare other lengths.
We analyse that segment with a short-time Fourier transform.
The transform has a 256-point (16\,ms) Hamming window and a 10\,ms hop.
Magnitudes are converted to decibels, with a dynamic range limited to 80\,dB.

\begin{itemize}
  \item \emph{Duration} is the length of the aligned segment.
  \item \emph{RMS amplitude} is the root-mean-square value of the whole filtered token.
  \item \emph{Burst amplitude} is found on the whole filtered token. We smooth the waveform with a Savitzky--Golay filter (window of five samples, polynomial order two). Peaks of the smoothed signal that exceed its standard deviation are candidate bursts. Burst amplitude is the largest value of the unsmoothed waveform at those peaks, or zero if there are none.
  \item \emph{Peak amplitude} is the largest value of the spectrogram in decibels.
  \item \emph{Peak frequency} is the frequency bin with the highest value of the time-averaged decibel spectrum.
  \item \emph{Dynamic amplitude} is the range, in decibels, between the largest and smallest values of the time--frequency representation.
  \item \emph{Spectral tilt} is the slope, in decibels per frequency bin, of a straight line fitted to the time-averaged spectrum.
  \item \emph{Spectral shape} is the standard deviation of the time-averaged spectrum across frequency bins.
\end{itemize}

RMS, peak and burst amplitude depend on recording level.
The other five features do not change when a constant gain is applied.
We do not normalise loudness.
Recordings are resampled to 16\,kHz and analysed at the level at which they were recorded or synthesised.
Differences in the three level-dependent features may therefore partly reflect level and not spectral content.

The earlier study used seven features \cite{bhyravajulla2025fricatives}: duration, dynamic amplitude, peak amplitude, peak frequency, RMS amplitude, spectral shape and spectral tilt.
The eight features here are those seven plus burst amplitude.

\subsubsection{Analysis}

We pair each synthetic token with a human token by the key defined in Section~\ref{subsec:common}.
Pairing means the two tokens in a pair share their linguistic content.
It does not remove all variation, so a remaining difference is not due to the voice alone.
A synthetic token with no human counterpart is left out.
This can happen, for example, when the aligner chooses a different pronunciation.

For each speaker--system pair we run a paired Wilcoxon signed-rank test \cite{wilcoxon1945individual}.
The test compares the two recordings within each stratum of context, manner class, voicing and feature.
The test does not assume that the differences are normally distributed.
It does assume that they are distributed symmetrically.

We correct for multiple tests with Holm's method \cite{holm1979simple}.
The correction is applied once per speaker--system pair, over all tests of that pair in all three contexts, at a significance level of 0.05.
It keeps the chance of any false positive within the pair below that level.
This family has about 260 tests per pair.
It includes the obstruent tests reported here and further tests that we do not report.

The effect size is Cliff's delta \cite{cliff1993dominance}.
It is the probability that a synthetic value is larger than a human value, minus the probability that it is smaller.
It ranges from $-1$ to $+1$, and a positive delta means that synthetic values are larger.
We compute it over all pairs of one synthetic and one human value in the stratum.
We do not restrict it to matched token pairs.
Its sign can therefore differ slightly from the sign of the paired test.
We report delta beside significance because, with hundreds of token pairs, even small differences become significant.

Tokens within an utterance are not independent, but the tests treat them as independent.
The p-values are therefore descriptive.

For comparison with the earlier study, we also analyse the eight LJSpeech systems with the unpaired Dunn's test \cite{dunn1964multiple} and Cliff's delta used there.
Each Dunn's test compares one system with its reference recordings, so the Benjamini--Hochberg adjustment \cite{benjamini1995controlling} has a single p-value to adjust and changes nothing.
We compare the two procedures by how often they agree on significance and by the correlation of their Cliff's delta values.

\subsubsection{Outcome measures}

The unit of analysis is a test: one speaker--system pair, context, manner class, voicing category and feature.
We call a test significant when it passes the Holm-corrected test within its pair.
Holm's correction does not span pairs.
The shares of significant tests are therefore descriptive summaries and not the result of one test.

We summarise a group of tests with three quantities.
The first is the share of significant tests.
The second is the mean absolute Cliff's delta.
The third is the sign of delta, which gives the direction of the difference.

Direction consistency is the share of significant tests of a feature whose sign matches the median sign of that corpus, system and feature.
The median comes from the same tests.
The share is therefore high whenever most tests share a sign.
It is weak evidence of consistency across speakers.

We base comparisons between groups on effect sizes and not on p-values.

\subsection{Experiment 2: statistical analysis of sonorants}
\label{subsec:p3-method}

\subsubsection{Setup}

Experiment 2 repeats the statistical analysis of Experiment 1 for sonorants.
The sonorants are nasals and approximants.
They are voiced and resonant.
Vocal-tract resonances shape their acoustics, and turbulent noise does not.

We therefore describe sonorants with features of resonance and not with the obstruent features.
This is a deliberate choice.
The obstruent features are designed for turbulent noise and do not suit sonorants, which are voiced and resonant.
We therefore chose features suited to sonorants.
The experiment does not test whether resonance cues are the ones that separate synthetic from human sonorants.
The nasal features follow a study of vowel nasality \citep{styler2017using}.
We apply them to nasal consonants as a starting point and did not test their suitability for consonants.
The classifiers of Experiment 3 use the same features, so that statistics and classifiers can be compared.

The statistics cover 188 speaker--system pairs.
These span all 31 speakers and all 16 system configurations: eight on VCTK and eight on LJSpeech.
We analyse sonorant tokens in the same way as obstruent tokens.

\subsubsection{Features}

\paragraph{Formant estimation}
Experiments 2 and 3 use the same class-specific formant features.
Each token is pre-emphasised with a coefficient of 0.97.
We model it with a linear predictive coding (LPC) filter of order 14 at 16\,kHz, estimated with the Burg method.
The order is a pragmatic choice for 16\,kHz speech and was not tuned.

We discard roots below 50\,Hz, with a bandwidth of 2000\,Hz or more, or with a non-positive bandwidth.
The remaining roots in the closed upper half of the unit circle are ordered by frequency, and real roots are included.
The $k$-th formant is the $k$-th lowest remaining root.
A root $r_k$ gives the formant frequency $F_k$ and the bandwidth $B_k$:
\begin{equation}
  F_k = \frac{f_s}{2\pi}\arg(r_k), \qquad B_k = -\frac{f_s}{\pi}\ln|r_k|,
\end{equation}
where $f_s=16$\,kHz.

We discard tokens whose formants fall outside plausible ranges for the phoneme class, as tracking failures.
We also discard tokens that cannot be analysed: unreadable audio, fewer than 24 samples, no usable root, approximants shorter than 42 samples, and approximants with fewer than two roots in the central third.
The retention in Table~\ref{tab:filter} counts all of these removals, and we did not measure how much of the loss comes from the ranges alone.
For nasals the first formant must lie between 150 and 900\,Hz, with a bandwidth between 20 and 500\,Hz.
For approximants the second formant must lie between 400 and 3500\,Hz and, if present, the third between 1000 and 4500\,Hz.
Table~\ref{tab:filter} gives the retention for each corpus, phoneme class and label.
On VCTK the filter removes about 18\% of nasal tokens and 0.3\% of approximant tokens, at similar rates for human and synthetic tokens.
On LJSpeech it removes human nasals slightly more often than synthetic nasals (4.4\% against 2.8\%).
Approximant removal is below 0.5\% in every group.
A difference in removal between human and synthetic tokens could bias the statistics and the classifier scores.
We did not measure the direction or size of any such bias.

\paragraph{Nasal features}
Eight features describe each nasal, following the acoustic account of nasality by Styler~\cite{styler2017using}.
\begin{itemize}
  \item The frequency and the bandwidth of the first formant.
  \item The amplitude difference between the first harmonic and the low-frequency nasal pole near 250--300\,Hz (A1--P0).
  \item The frequency and the depth of the spectral antiformant. It is the minimum of the time-averaged spectrum within a band specific to the place of articulation: 750--1250\,Hz for /m/, 1450--2200\,Hz for /n/ and 2800--4000\,Hz for /\textipa{N}/. Depth is the mean level of the neighbouring bins minus the level at the minimum.
  \item The ratio of the energy below 500\,Hz to the energy up to 8\,kHz, computed from the decibel spectrum after converting it back to a linear scale, with a floor 80\,dB below its peak.
  \item The spectral tilt between 200 and 2000\,Hz.
  \item The duration of the token.
\end{itemize}
The first harmonic is located within 50\,Hz of the fundamental frequency of the token.
That frequency is the median over frames of the YIN pitch tracker \cite{decheveigne2002yin}, restricted to 60--400\,Hz.
We use 120\,Hz when the tracker finds no pitch.
The time-averaged spectra use a 512-point Hamming-window short-time Fourier transform with a hop of 10\,ms (160 samples).
For short tokens the transform length is reduced to the largest power of two that fits, and the hop shrinks with it.
The conversion from frequency to bin still assumes a 512-point transform, which affects A1--P0 and the antiformant features for tokens shorter than 512 samples (32\,ms), about 10\% of the kept nasal tokens.

\paragraph{Approximant features}
Nine features describe each approximant, motivated by work on formant dynamics in spoofed speech \cite{liu2024formant}.
We divide the token into three equal parts and estimate the formants in each part.
\begin{itemize}
  \item The values of the second to fifth formants in the central third of the token (four features).
  \item The changes in the second and third formants between the first and last thirds of the token (two features).
  \item The bandwidth of the third formant.
  \item The distance between the second and third formants.
  \item The duration of the token.
\end{itemize}

\subsubsection{Analysis}

The procedure of Experiment 1 is applied.
We pair synthetic and human tokens by the key defined in Section~\ref{subsec:common}, within each context.
Only tokens that pass the formant validity filter are included.
A paired Wilcoxon signed-rank test compares the two within each stratum of context, phoneme class and feature.
As in Experiment 1, one such comparison is one test.

Holm's correction is applied once per speaker--system pair, over the sonorant tests of that pair only.
There are up to 51 such tests: seventeen features (eight nasal and nine approximant) in each of the CV, VC and isolated contexts.
This family is smaller than the one in Experiment 1, so significance in the two experiments is not corrected on the same footing.
A test is skipped when fewer than two token pairs exist or all paired differences are zero.
The isolated tests enter the correction and are reported separately.
The effect size is Cliff's delta, computed as in Experiment 1.

\subsubsection{Outcome measures}

We report the share of significant tests, the mean absolute Cliff's delta and the sign of delta, as in Experiment 1.
The isolated context has a median of only 10 matched pairs per test.
We therefore report it separately and exclude it from the other figures.

\subsection{Experiment 3: classifiers on obstruents and sonorants}
\label{subsec:p2-method}

\subsubsection{Setup}

Experiment 3 trains classifiers on the obstruent tokens of Experiment 1 and the sonorant tokens of Experiment 2.
A classifier shows how well a set of features separates synthetic from human tokens that it has not seen.
Prior statistical work points to fricatives as the obstruents that separate synthetic from human speech most clearly \cite{bhyravajulla2025fricatives}.
We therefore train classifiers for fricatives, stops and affricates with the same features and the same protocol.
If fricatives separate the classes most clearly, fricative classifiers should score highest.
For sonorants, we train classifiers for nasals and approximants with the formant features of Experiment 2.

\paragraph{Obstruents}
The unit of analysis is a classifier cell.
A cell holds the tokens of one speaker, one TTS system, one voicing category (voiced, voiceless or both) and one context (CV, VC, isolated or all).
The synthetic class is the tokens from the TTS system.
The human class is all human tokens of the same speaker in that voicing category and context.

Each manner class has up to 2{,}256 cells: 188 speaker--system pairs, three voicing categories and four contexts.
A cell is kept if each class has at least 10 tokens and each side of the split has at least 5 tokens of each class.
All 2{,}256 fricative cells and all 2{,}256 stop cells are kept.
Only 1{,}667 affricate cells are kept, because affricates are rare.
The test sets have a median of 379 tokens for fricatives (range 19 to 11{,}271) and 312.5 for stops (53 to 11{,}885).
The affricate test sets are small, with a median of 24 tokens (range 10 to 565).

For affricates, voiced means /\textipa{dZ}/ only and voiceless means /\textipa{tS}/ only.

The cells are nested.
A cell with voicing ``both'' or context ``all'' contains the tokens of the cells with a single voicing category or context of the same speaker and system.
Means over groups of cells therefore count some tokens more than once.
The cells of one speaker are also not independent, because they share that speaker's human tokens.
We report means over cells.
Intervals are 95\% percentile intervals from 5{,}000 resamples of speakers with replacement, which keeps all cells of a speaker together.
The same procedure gives the intervals for differences between system groups on shared speakers, where the difference is conventional minus cloning.

\paragraph{Sonorants}
The sonorant classifiers use the same speakers and systems as Experiment 2.
Nasals and approximants are voiced, so every cell uses the voiced category.
A cell is one speaker, system, phoneme class and context (CV, VC or isolated).
There is no pooled ``all'' context cell.
There are 422 nasal cells and 447 approximant cells.

A cell is kept if it holds at least 30 tokens in total, with at least 20 for training and 10 for testing after the split.
We check this again after tokens with missing feature values are dropped.
The \auc is defined in every kept cell.
Table~\ref{tab:minsize} collects the minimum-size rules of all experiments.

There are 188 speaker--system pairs and three contexts, which gives 564 candidate cells per class.
The rule keeps 422 of the 564 nasal cells and 447 of the 564 approximant cells.
LJSpeech supplies 24 cells per class (eight systems and three contexts), and VCTK supplies the rest.
The median test set holds 124 tokens for nasals and 284 for approximants.
For the neural network, 421 of the 422 nasal cells were fitted.
In the remaining cell the fit failed, because batch normalisation needs more than one sample in a batch.

\subsubsection{Features}

\paragraph{Obstruents}
Six of the Experiment 1 features are used: duration, RMS amplitude, peak frequency, dynamic amplitude, spectral tilt and spectral shape.
They are computed as in Experiment 1.
Duration is the interval length of the aligned token, available at a resolution of about 10\,ms.

Two features are left out.
Peak amplitude is left out because it depends on recording level.
Burst amplitude is left out because it is meant to capture the release burst of stops and affricates.
Experiment 1 computes it for all obstruents as the largest transient in the token, so its meaning for a fricative is not established.

RMS amplitude depends on recording level in the same way as peak amplitude, but we keep it.
The subset therefore does not follow a single selection rule.

The human class of a cell is all human tokens of that speaker, including those from sentences that the system did not synthesise.
The two classes need not share the same sentences.

\paragraph{Sonorants}
The sonorant classifiers use the features of Experiment 2: eight for nasals and nine for approximants, each set including duration.

\paragraph{Class-neutral baseline}
The obstruent features were developed for fricatives, and the formant features for sonorants.
To compare all five classes with one set of features, we add a class-neutral baseline.
For each token we compute mel-frequency cepstral coefficients (MFCC) and linear-frequency cepstral coefficients (LFCC).
The token is pre-emphasised with a coefficient of 0.97 and cut into 25\,ms Hamming frames with a 10\,ms hop.
A 512-point Fourier transform gives the power spectrum.
Forty triangular filters cover 0 to 8\,kHz, on the mel scale for MFCC and on a linear scale for LFCC.
We take the logarithm of the filter energies and apply a discrete cosine transform.
We keep coefficients 1 to 12 and drop the energy term, so the features do not depend on level.
For each coefficient we take the mean and the standard deviation over the frames of the token.
This gives 24 features for each representation.
As for the class-specific features, we add duration, which gives 25 features.
Frames are centred with reflection padding, and tokens shorter than one frame are zero-padded.
The mel filters follow the HTK mel scale.

\subsubsection{Analysis}

\paragraph{Obstruent classifiers}
We train a gradient-boosted decision tree classifier (XGBoost \cite{chen2016xgboost}) separately for each cell.
Training per cell fixes the speaker and the system.
Each score therefore shows how detectable one system is for one voice.

The classifier uses 300 trees of maximum depth 4, a learning rate of 0.1 and a row subsampling rate of 0.8.
The other parameters are the library defaults.
The weight of the synthetic class is set to the ratio of human to synthetic training tokens in the cell.
The random seed is fixed at 42 for the split and the classifier.
The hyperparameters are fixed values, and there is no validation split.

Each cell is divided 75\%/25\% into training and test tokens, once.
Each score therefore comes from a single split.
All tokens from one utterance fall on the same side of the split.
This keeps tokens of the same utterance out of both sets.
Because whole utterances are assigned to a side, about 25\% of utterances fall in the test set, and not exactly 25\% of tokens.

To compare manner classes, we pair their cells by speaker, system, voicing category and context.
We take the difference in \auc for each pair of cells.
We resample speakers to obtain intervals for the mean difference.

\paragraph{Recording level}
We do not normalise loudness, and RMS amplitude depends on recording level.
To test its influence, we repeat the analysis without RMS amplitude.
The remaining five features do not change when a constant gain is applied.
All other settings stay the same.

\paragraph{Matched tokens}
In the analysis above, the human class of a cell holds all human tokens of the speaker.
The two classes therefore need not share sentences or phonemes.
To test whether this matters, we repeat the analysis on matched tokens.
A cell keeps a human token only if the system synthesised the same token, that is, the same sentence, word, word position and phoneme position.
It keeps a synthetic token only if the speaker has a human counterpart.
Each class then holds the same linguistic tokens.
All other settings stay the same.
We apply the same matching to the sonorant classifiers, for XGBoost only.

\paragraph{Per-phoneme classifiers}
We also train a classifier for each fricative and stop phoneme separately.
A cell is one speaker, system and context for one phoneme.
The voicing of a phoneme is fixed, so no voicing categories are used.
The human class is the speaker's human tokens of that phoneme, and the synthetic class is the system's tokens of that phoneme.
The features, split, size rules and settings are those of the main analysis.

\paragraph{Feature contributions}
We measure the contributions of the features for fricatives with SHAP values \cite{lundberg2017unified}.
SHAP values are computed on the test tokens of each cell.
For each cell we take the mean signed SHAP value of each feature over those tokens.
We then average the absolute value of that mean over cells.
Positive and negative contributions within a cell therefore cancel before the absolute value is taken.

\paragraph{Unseen cells}
We test generalisation to unseen cells for all five phoneme classes.
This asks whether the cues learned for one voice and system carry over to other voices and systems.
We train an XGBoost classifier with the settings above on all tokens of one cell, with no held-out split.
We then evaluate it on all tokens of a different cell.
Both cells must have at least 30 tokens of each class.
This minimum applies to this analysis only.
We use only cells that pool all voicing categories and all contexts.
Each class uses its own features.
Sonorants have no pooled context cell in the main analysis, so a sonorant cell pools the CV, VC and isolated contexts of one speaker and system.

Three conditions are defined by what differs between the training cell and the test cell.
\begin{itemize}
  \item \textbf{Cross-system}: same speaker, different TTS system.
  \item \textbf{Cross-speaker}: same TTS system, different speaker.
  \item \textbf{Cross-both}: different speaker and different TTS system.
\end{itemize}
The within-cell reference is the mean \auc of the main analysis, from held-out tokens of the same cell, over the VCTK cells that pool all voicing categories and contexts.
We evaluate each condition within VCTK and, where cells exist, within LJSpeech.
Pairs that cross the two corpora are reported separately.

Systems are matched by identical identifier.
Across corpora only VITS has the same identifier.
The other LJSpeech identifiers name the vocoder.
The same architecture trained on the two corpora therefore counts as a different system.
LJSpeech has one speaker, so the cross-speaker condition within LJSpeech is empty.
Pairs across corpora with the same system and a different speaker are cross-speaker pairs.
There are 30 pairs from the LJSpeech VITS cell to the 30 VCTK VITS cells, and 30 in the opposite direction.
This gives 60 pairs and 31 test cells.
All other pairs across corpora are cross-both pairs.
Eight LJSpeech cells against 180 VCTK cells, less the 30 same-system combinations, give 1{,}410 pairs in each direction.
This gives 2{,}820 pairs and 188 test cells.

Each test cell appears in many pairs.
We therefore average the pair scores within each test cell before computing summary statistics.
In cross-system pairs the human tokens of the training and test cell come from the same speaker, so they are shared between the two cells.
Recall values are means over the pairs of a condition, at a probability threshold of 0.5.

To help interpret the cross-corpus results, we also score each feature alone.
For each corpus we pool every fricative token of all systems and speakers.
One feature at a time is the score, with the synthetic class as the positive class.
An \auc below 0.5 then means that the feature is lower in synthetic than in human tokens.
We compute the same score between the human tokens of the two corpora, with LJSpeech as the positive class.

\paragraph{Class-neutral classifiers}
We train the same per-cell XGBoost classifiers on the 24 MFCC features plus duration and on the 24 LFCC features plus duration, for all five classes.
Fifteen tokens have no segment audio and are left out.
The grid, split, size rules and settings are those of the main analysis.
Sonorant cells are the voiced cells of the CV, VC and isolated contexts, as in the main sonorant analysis.
To compare fricatives with another class, we pair cells by speaker, system, voicing category and context.
We take the difference in \auc for each pair of cells, with speaker-cluster bootstrap intervals as above.
Sonorant cells are therefore paired with the voiced fricative cells of the same context.

\paragraph{Sonorant classifiers}
We train four classifiers on identical cells and splits.
We use the 75\%/25\% utterance-grouped protocol of the obstruent classifiers.
Cells with fewer than four utterances are split sequentially: the first 75\% of the tokens in the order of the dataset are used for training and the rest for testing.
In the reported analyses no sonorant cell was split sequentially.
The human class of a cell is all human tokens of the same speaker, phoneme class and context, as for obstruents.

The four classifiers are the following.
\begin{itemize}
  \item XGBoost, with the settings used for obstruents.
  \item $\ell_2$-regularised logistic regression, with regularisation strength 1.
  \item A linear support vector machine \cite{cortes1995support} with regularisation strength 1. It scores a token by its signed distance to the decision boundary, and its balanced accuracy uses a score of zero as the threshold.
  \item A feed-forward neural network with three blocks of width 8, batch normalisation \cite{ioffe2015batch} and GELU activations \cite{hendrycks2016gelu}. It is trained with a class-weighted loss for 100 epochs with the Adam optimiser \cite{kingma2014adam}, a learning rate of $10^{-3}$ and batches of up to 512 tokens, without early stopping.
\end{itemize}
The linear models use standardised features and balanced class weights.
The seed is fixed at 42 for all classifiers.
We use four classifiers so that a difference between feature sets can be checked against more than one model family.
This comparison does not by itself show whether a limit lies in the features or in the model.
We also analyse approximant cells separately for each phoneme, with the same size rule, which gives one cell per speaker, system, context and phoneme.

We compute SHAP values of the XGBoost classifiers on the test tokens of each cell.
We take the absolute value per token and average it over tokens and then over cells.
Software versions are not reported (Section~\ref{sec:limitations}).

\paragraph{Transfer between corpora}
VITS is the one system with the same identifier on both corpora, so we test transfer between corpora for VITS.
We train an XGBoost classifier on all tokens of one speaker--system cell and test it on all tokens of another cell.
There are three directions.
\begin{itemize}
  \item From each VCTK VITS speaker to the LJSpeech VITS speaker.
  \item From the LJSpeech VITS speaker to each VCTK VITS speaker.
  \item From one VCTK VITS speaker to another VCTK VITS speaker, as a reference within one corpus.
\end{itemize}
Each cell must have at least 30 tokens in each class.
We use the CV and VC contexts only, because isolated cells are too small.
Class weights and settings are as above.

\subsubsection{Outcome measures}

Four metrics score each classifier on held-out tokens.
\begin{itemize}
  \item The area under the ROC curve (\auc) is the probability that a randomly chosen synthetic token receives a higher score than a randomly chosen human token. A value of 0.5 is chance and 1 is perfect. It does not depend on a threshold.
  \item Average precision (\ap) is the area under the precision--recall curve. It depends on the share of synthetic tokens in the test set.
  \item The equal error rate (\eer) is the error rate at the threshold where the share of synthetic tokens called human equals the share of human tokens called synthetic. We take it at the point of the ROC curve where the two rates are closest. Lower is better.
  \item Balanced accuracy (\balacc) is the mean of recall on the synthetic class and recall on the human class, at a threshold of 0.5.
\end{itemize}
\auc, \eer and balanced accuracy do not depend on the share of each class.
Average precision does, so we read it with that share in mind.
Reported values are means over the cells that have a defined \auc.
Intervals for the means are speaker-cluster bootstrap intervals, as described above.
The sonorant headline in Table~\ref{tab:p3-headline} averages over all contexts, including the isolated context.
The XGBoost means for the CV and VC contexts alone are in Table~\ref{tab:p3-context}.
We do not report a pooled CV and VC mean for the other classifiers.

\section{Results}
\label{sec:results}

In the statistics of Experiments 1 and 2, a test is significant when it passes the Holm correction within its speaker--system pair.
The shares of significant tests are descriptive.
Tokens within an utterance are treated as independent, and no correction spans pairs.
We base comparisons on effect sizes.
Values are rounded to two decimals unless stated otherwise.

\subsection{Experiment 1: statistical analysis of obstruents}
\label{subsec:p1-results}

\subsubsection{Overall}

About half of the obstruent tests are significant (0.51 pooled; Table~\ref{tab:p1-system}).
The share differs by manner class (Table~\ref{tab:p1-manner}).
Over the eight features, 0.63 of voiced and 0.75 of voiceless fricative tests are significant.
For stops the shares are 0.58 (voiced) and 0.72 (voiceless).
For affricates, they are 0.15 (voiced) and 0.17 (voiceless).
The median affricate test has 23 token pairs, so we report affricate effect sizes without interpretation.

\begin{figure}[!htbp]
\centering
\includegraphics[width=\textwidth]{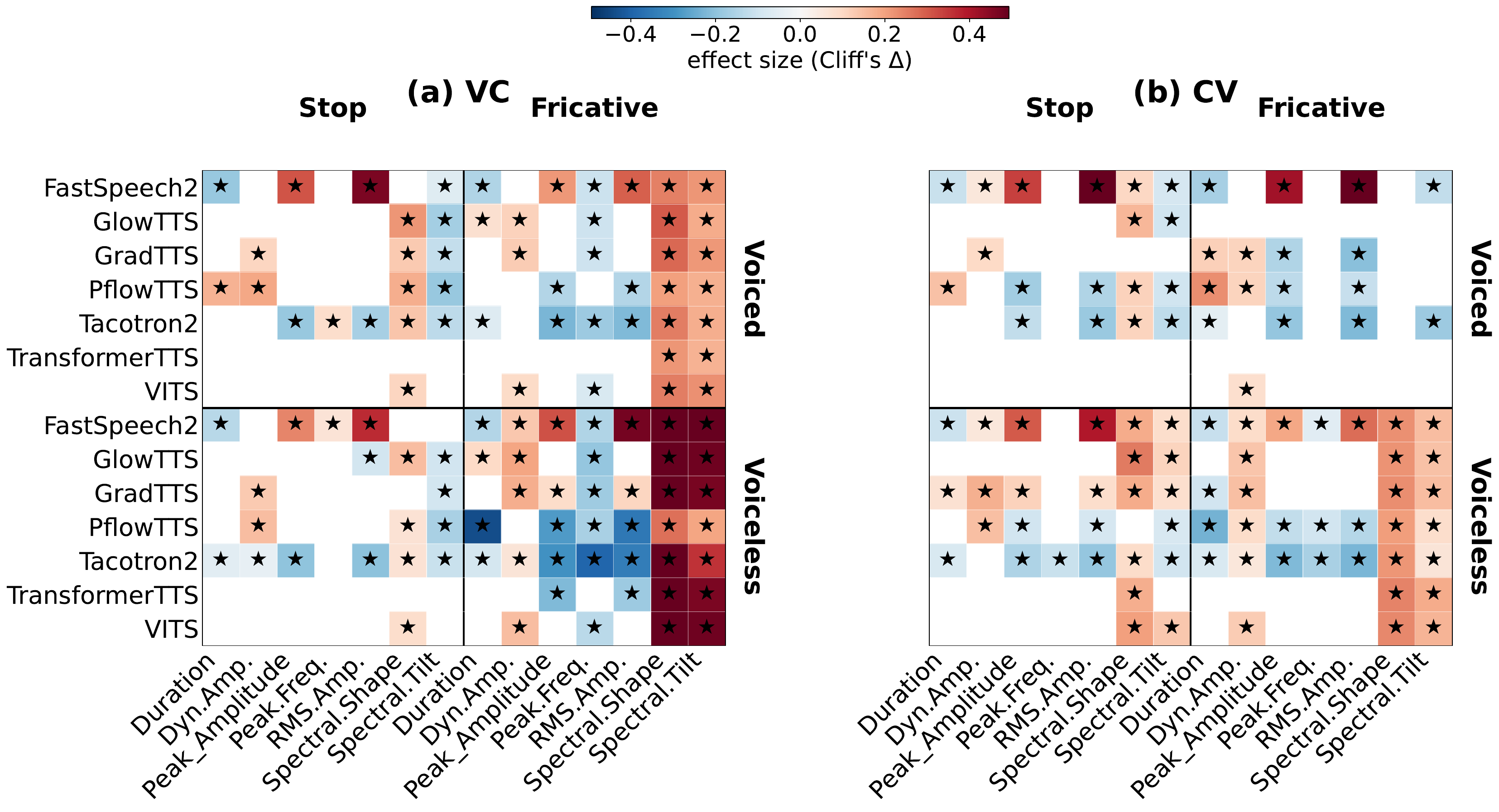}
\caption{Experiment 1: effect size (Cliff's delta) for stops and fricatives, by system, voicing and feature, in (a) VC and (b) CV context. Each system is compared with its LJSpeech reference recordings. DiffGAN-TTS and burst amplitude are not shown. Red means a higher value in synthetic speech and blue a lower value. A star marks a significant difference (Dunn's post-hoc test, $p<0.001$), and blank cells are not significant. The test is unpaired, so it differs from the paired tests in the tables. Fricative columns show the largest and most consistent effects, in spectral shape and spectral tilt.}
\label{fig:p1-heatmap}
\end{figure}

\subsubsection{By context}

Significance is similar in the CV and VC contexts and lower in the isolated context (Table~\ref{tab:p1-context}).
The mean absolute delta is 0.24 in CV, 0.23 in VC and 0.27 in the isolated context.
The isolated context therefore has the lowest share of significant tests and the largest mean absolute delta.
Token-pair counts per context are not tabulated, and we did not test the reason for the lower isolated share.

\subsubsection{By phoneme class and voicing}

Voiceless tests are significant more often than voiced tests in every manner class (Table~\ref{tab:p1-manner}).
For fricatives, 0.75 of voiceless tests and 0.63 of voiced tests are significant.
Figure~\ref{fig:p1-heatmap} shows the effect sizes behind these tests for stops and fricatives, by system, voicing and feature.
The fricative columns for spectral shape and spectral tilt show the largest and most consistent effects.

\begin{table}[!htbp]
\centering
\caption{Experiment 1: significant tests (Holm, $p<0.05$) and mean absolute Cliff's delta by manner class and voicing, over eight features, VCTK and LJSpeech pooled. The last two rows restrict the fricative tests to the six features used in Experiment 3, and their test counts are not given. The voiceless affricate group has fewer tests than the other groups. The median affricate test has 23 token pairs.}
\label{tab:p1-manner}
\begin{tabular}{llccc}
\toprule
Manner & Voicing & Tests & Significant & Mean $|\delta|$ \\
\midrule
Fricative & Voiced    & 4{,}512 & 0.63 & 0.23 \\
Fricative & Voiceless & 4{,}512 & 0.75 & 0.21 \\
Stop      & Voiced    & 4{,}512 & 0.58 & 0.18 \\
Stop      & Voiceless & 4{,}512 & 0.72 & 0.18 \\
Affricate & Voiced    & 4{,}504 & 0.15 & 0.33 \\
Affricate & Voiceless & 3{,}973 & 0.17 & 0.37 \\
\midrule
Fricative, six features & Voiced & -- & 0.61 & 0.22 \\
Fricative, six features & Voiceless & -- & 0.73 & 0.20 \\
\bottomrule
\end{tabular}
\end{table}

Significance and effect size point in different directions for fricative voicing.
On the six features used in Experiment 3, voiceless fricatives are significant more often than voiced ones.
Voiced fricatives have the larger mean absolute delta (last two rows of Table~\ref{tab:p1-manner}).

\subsubsection{By system}

Voice-cloning systems differ from human speech less often than conventional systems (Table~\ref{tab:p1-system}).
Even so, more than half of the fricative tests remain significant for the cloning systems, and their mean absolute delta for fricatives is smaller (Table~\ref{tab:p1-shared}).
The two groups cover different speakers, 30 for the conventional systems and 10 for the cloning systems.
We therefore also compare the groups on the 10 speakers they share.
On those speakers the conventional systems have the larger share of significant tests and the larger mean absolute delta.
This holds for all obstruents and for fricatives alone (Table~\ref{tab:p1-shared}).

\begin{table}[!htbp]
\centering
\caption{Experiment 1: significant tests and mean absolute Cliff's delta by system group, over eight features and all obstruent classes. Fricative-only figures for the cloning group are in Table~\ref{tab:p1-shared}. The last row pools the three groups, and its values are test-weighted means of the rows above.}
\label{tab:p1-system}
\begin{tabular}{lccc}
\toprule
System group & Tests & Significant & Mean $|\delta|$ \\
\midrule
Conventional VCTK (5 systems, 30 speakers)  & 21{,}150 & 0.53 & 0.26 \\
Voice cloning (3 systems, 10 speakers)      & 4{,}231  & 0.39 & 0.19 \\
LJSpeech (8 systems)                        & 1{,}144  & 0.51 & 0.19 \\
\midrule
All three groups                            & 26{,}525 & 0.51 & 0.25 \\
\bottomrule
\end{tabular}
\end{table}

Among the VCTK systems, DiffGAN-TTS differs most from human speech over all obstruent classes (Table~\ref{tab:p1-persystem}).
Its mean absolute delta is 0.32.
The order of the remaining systems by mean absolute delta is Tacotron~2, FastSpeech~2, Transformer~TTS, FishSpeech, CosyVoice~2, Qwen3-TTS and VITS.

\subsubsection{By feature}

Features differ in how often they separate synthetic from human speech (Table~\ref{tab:p1-feature}).
RMS amplitude, spectral shape and burst amplitude are significant most often.
Peak frequency is significant least often (0.21).

Spectral tilt is the only feature that is higher in synthetic speech in the median significant fricative test.
Dynamic amplitude and spectral shape are lower in synthetic speech.
Synthetic segments are slightly shorter.
These directions pool VCTK and LJSpeech, and VCTK dominates them: it supplies 25{,}381 of the 26{,}525 tests.
On LJSpeech the single-feature check of Table~\ref{tab:p2-single} shows the opposite direction for dynamic amplitude and spectral shape.
Table~\ref{tab:p1-direction} gives the share of significant fricative tests in which tilt is higher in synthetic speech.

\subsubsection{Consistency and generalisation}

Among significant fricative tests, the sign of delta mostly matches the median sign of the same corpus, system and feature (Table~\ref{tab:p1-direction}).
This share is high by construction whenever one sign dominates, so it is weak evidence of consistency across speakers.

The paired analysis and the unpaired Dunn's test of the earlier work agree on significance in 0.88 of the 1{,}144 LJSpeech obstruent tests.
The Cliff's delta values of the two procedures correlate at 1.00 (0.999 before rounding; Table~\ref{tab:p1-dunn}).

\subsection{Experiment 2: statistical analysis of sonorants}
\label{subsec:p3-results}

\subsubsection{Overall}

The isolated context has a median of 10 matched pairs per test, and 0.005 of its tests are significant.
We exclude it from the figures below.
In the CV and VC contexts, 6{,}392 tests cover 188 speaker--system pairs, with a median of 536 matched pairs per test (Table~\ref{tab:p3-pairs}).
Of these tests, 0.41 are significant, with a mean absolute Cliff's delta of 0.10.
Nasals (0.40, $|\delta|$ 0.12) and approximants (0.43, 0.09) differ from human speech in a similar share of tests.

\subsubsection{By context}

Significance differs by context and phoneme class (Table~\ref{tab:p3-stat-context}).
Nasals are significant more often in VC (0.49) than in CV (0.30).
Approximants are significant slightly more often in CV (0.45) than in VC (0.41).
Nasal CV tests have a median of 185.5 matched pairs, against 693.5 in VC (Table~\ref{tab:p3-pairs}).

\subsubsection{By phoneme class}

Nasals have the larger effects (mean $|\delta|$ 0.12 against 0.09).
Approximants are significant slightly more often (0.43 against 0.40).

\subsubsection{By system}

Voice-cloning systems differ from human speech more often than conventional systems (Table~\ref{tab:p3-stat-system}).
The groups cover different speakers, so Table~\ref{tab:p3-shared} also gives the figures on the ten speakers that both groups share.
On those speakers the cloning systems have the larger share of significant tests and the larger mean absolute delta, for nasals and for all sonorants.
For approximants the interval of the difference in mean absolute delta includes zero.
This is the opposite pattern to that of the obstruents in Experiment 1.

Among the VCTK systems, DiffGAN-TTS has the highest share of significant tests and the largest mean absolute delta (Table~\ref{tab:p3-stat-persystem}).
VITS has the lowest, for nasals and for approximants.
LJSpeech VITS also has a low share: 0.25 of its 16 nasal tests and 0.33 of its 18 approximant tests are significant, against 0.45 for the LJSpeech systems as a whole.

\begin{table}[!htbp]
\centering
\caption{Experiment 2: significant tests and mean absolute Cliff's delta by system group, CV and VC contexts.}
\label{tab:p3-stat-system}
\begin{tabular}{lccc}
\toprule
System group & Tests & Significant & Mean $|\delta|$ \\
\midrule
Conventional VCTK (5 systems) & 5{,}100 & 0.39 & 0.10 \\
Voice cloning (3 systems) & 1{,}020 & 0.50 & 0.12 \\
LJSpeech (8 systems) & 272 & 0.45 & 0.05 \\
\bottomrule
\end{tabular}
\end{table}

\subsubsection{By feature}

Significance differs widely across the formant features (Table~\ref{tab:p3-stat-feature}).
For approximants, the bandwidth of F3 is significant most often, and the F2 and F3 slopes least often.
For nasals, the spectral tilt between 200 and 2000\,Hz and the low-band energy ratio are significant most often, and the antiformant features least often.
Duration is significant in 0.51 of the nasal tests and 0.48 of the approximant tests.

\subsubsection{Consistency}

Among the significant tests, the direction of an effect is not the same for all features.
The bandwidth features and the nasal low-band energy ratio are higher in synthetic speech in most significant tests.
The antiformant features are lower in about two thirds of them.
F2 mean has no consistent direction (Table~\ref{tab:p3-direction}).
These sonorant effects cannot be compared directly with the obstruent directions of Experiment 1, because the features differ.

\subsection{Experiment 3: classifiers on obstruents and sonorants}
\label{subsec:p2-results}

\subsubsection{Overall}

\paragraph{Obstruents}
Fricatives are classified best (Table~\ref{tab:p2-headline}).
The six features separate synthetic from human fricatives with a mean \auc of 0.95 over 2{,}256 cells (95\% interval 0.94 to 0.95).
Affricates come next, with a mean \auc of 0.85 over 1{,}667 cells (0.84 to 0.86).
Stops are lowest, with 0.83 over 2{,}256 cells (0.81 to 0.84).
The interval of fricatives does not overlap with those of the other two classes.
The intervals of affricates and stops overlap.
The same order holds for average precision, equal error rate and balanced accuracy.

\begin{table}[!htbp]
\centering
\caption{Experiment 3: mean performance over all cells of each manner class. The interval is a 95\% speaker-cluster bootstrap interval for the mean \auc. The last column gives the median and range of the number of test tokens in a cell.}
\label{tab:p2-headline}
\begin{tabular}{lccccccc}
\toprule
Manner class & Cells & \auc & 95\% interval & \ap & \eer & \balacc & Test tokens \\
\midrule
Fricatives & 2{,}256 & 0.95 & 0.94--0.95 & 0.93 & 0.12 & 0.88 & 379 (19--11{,}271) \\
Stops & 2{,}256 & 0.83 & 0.81--0.84 & 0.81 & 0.25 & 0.75 & 312.5 (53--11{,}885) \\
Affricates & 1{,}667 & 0.85 & 0.84--0.86 & 0.85 & 0.21 & 0.78 & 24 (10--565) \\
\bottomrule
\end{tabular}
\end{table}

The comparison above uses different sets of cells, so we also compare the classes on the 1{,}667 cells that all three share (Table~\ref{tab:p2-matched}).
On these cells the mean \auc is 0.95 for fricatives, 0.85 for affricates and 0.83 for stops.
Fricatives have a higher \auc than affricates in 0.82 of the cells, and a higher \auc than stops in 0.99 of the cells.
The mean gap is 0.09 between fricatives and affricates and 0.11 between fricatives and stops.
Affricates are above stops by 0.02, and in 0.64 of the cells.

\begin{table}[!htbp]
\centering
\caption{Experiment 3: paired comparison of manner classes on the 1{,}667 cells that all three classes share (same speaker, system, voicing and context). The gap is the mean difference in \auc, with a 95\% speaker-cluster bootstrap interval. The last column is the share of cells in which the first class has the higher \auc.}
\label{tab:p2-matched}
\begin{tabular}{lccc}
\toprule
Comparison & Mean gap & 95\% interval & Share of cells \\
\midrule
Fricative minus affricate & 0.09 & 0.08--0.11 & 0.82 \\
Fricative minus stop & 0.11 & 0.10--0.13 & 0.99 \\
Affricate minus stop & 0.02 & 0.01--0.03 & 0.64 \\
\bottomrule
\end{tabular}
\end{table}

Two cautions apply to the affricates.
The affricate test sets are small, with a median of 24 tokens, so single affricate cells are noisy.
The 589 affricate cells that fail the size rule have too few tokens in one class or on one side of the split.
The affricate mean therefore describes the better-populated cells and may be too high.

Matching the tokens changes the scores very little (Table~\ref{tab:matched-tokens}).
In the main analysis the two classes need not share sentences or phonemes.
In the matched analysis each human token has a synthetic counterpart of the same sentence, word, word position and phoneme position, and the reverse.
The mean \auc is 0.95 for fricatives, 0.83 for stops and 0.85 for affricates, against 0.95, 0.83 and 0.85 in the main analysis.
The mean change per cell is at most 0.001 in absolute value for every obstruent class.
For stops, it is $-0.001$, and its interval includes zero.
The order of the classes stays the same.
On the 1{,}652 matched-token cells that all three classes share, the mean \auc is 0.95 for fricatives, 0.85 for affricates and 0.83 for stops.
The fricative lead is therefore not explained by differences in sentences or phonemes between the human and synthetic tokens.

\paragraph{Sonorants}
The formant features detect synthetic nasals and approximants above chance (Table~\ref{tab:p3-headline}).
XGBoost has the highest mean \auc for both classes: 0.71 for nasals (95\% interval 0.69 to 0.73) and 0.76 for approximants (0.75 to 0.77).
For nasals the four classifiers lie within 0.03 \auc of one another, and their intervals overlap.
For approximants the neural network (0.70; 0.68 to 0.71) is below XGBoost and above the linear models (0.64 for both).
For approximants XGBoost exceeds the linear models by about 0.12.
These values are lower than the fricative value (0.95), but the feature sets differ.
Table~\ref{tab:p3-context} gives the XGBoost means for the CV and VC contexts alone, which exclude the isolated context.
Matching the tokens leaves the sonorant scores unchanged as well (Table~\ref{tab:matched-tokens}).
Over the cells that remain, the mean \auc is 0.71 for nasals and 0.78 for approximants in both analyses.
These means cover only the cells that remain, so they can differ from the headline values.
The mean change per cell is at most 0.006 in size for the sonorants.
Matching lowers the number of cells from 422 to 398 for nasals and from 447 to 387 for approximants, because some cells fall under the size rule.

\begin{table}[!htbp]
\centering
\caption{Experiment 3: mean classifier performance by phoneme class, over all cells including the isolated context (422 nasal and 447 approximant cells; 421 for the neural network on nasals). The \auc is defined in every cell. The interval is a 95\% speaker-cluster bootstrap interval for the mean \auc.}
\label{tab:p3-headline}
\begin{tabular}{llccccc}
\toprule
Class & Classifier & \auc & 95\% interval & \ap & \eer & \balacc \\
\midrule
Nasal & XGBoost & \textbf{0.71} & 0.69--0.73 & 0.69 & 0.34 & 0.66 \\
 & Logistic regression & 0.69 & 0.67--0.71 & 0.65 & 0.36 & 0.64 \\
 & Linear SVM & 0.69 & 0.67--0.71 & 0.65 & 0.36 & 0.64 \\
 & Neural network & 0.69 & 0.67--0.71 & 0.66 & 0.36 & 0.63 \\
\midrule
Approximant & XGBoost & \textbf{0.76} & 0.75--0.77 & 0.74 & 0.30 & 0.70 \\
 & Logistic regression & 0.64 & 0.63--0.66 & 0.60 & 0.39 & 0.61 \\
 & Linear SVM & 0.64 & 0.63--0.66 & 0.60 & 0.39 & 0.61 \\
 & Neural network & 0.70 & 0.68--0.71 & 0.67 & 0.35 & 0.65 \\
\bottomrule
\end{tabular}
\end{table}

\paragraph{Class-neutral baseline}
We repeated the comparison with class-neutral features (Table~\ref{tab:cn}).
Both baselines separate the sonorants much better than the formant features do.
With LFCC the mean \auc is 0.98 for fricatives, 0.96 for affricates, 0.95 for stops and nasals, and 0.92 for approximants.
With MFCC it is 0.91 for fricatives, 0.88 for nasals, 0.87 for affricates and stops, and 0.83 for approximants.
Fricatives rank first in the mean under both representations, although with MFCC they cannot be told apart from nasals.
The paired comparison shows how large the lead is (Table~\ref{tab:cn-gaps}).
With LFCC, fricatives are ahead of every other class, and every interval excludes zero.
The gap is 0.020 for affricates, 0.024 for nasals, 0.028 for stops and 0.055 for approximants.
For affricates the gap is small, and fricatives are ahead in only 0.54 of the shared cells.
With MFCC, fricatives are ahead of stops, affricates and approximants, by 0.036 to 0.053.
They are tied with nasals: the gap is 0.004, with an interval of $-0.011$ to 0.022.
Fricatives have the highest \auc in 14 of 16 system configurations with LFCC, and in 10 with MFCC.
With the class-specific features they have the highest \auc in 15 of 16.
The lead of fricatives is therefore clear with the class-specific features and with LFCC, and it is not clear with MFCC.
For sonorants, the formant features were weaker than the class-neutral features.

\begin{table}[!htbp]
\centering
\caption{Experiment 3: class-neutral baseline. Mean \auc over all cells of each class with the class-specific features (as above), with MFCC summary statistics, and with LFCC summary statistics. The intervals are 95\% speaker-cluster bootstrap intervals. The last four rows give the equal error rate and balanced accuracy of the two baselines. Each baseline uses 24 cepstral features and duration per token, the same for every class.}
\label{tab:cn}
\resizebox{\ifdim\width>\textwidth\textwidth\else\width\fi}{!}{%
\begin{tabular}{lccccc}
\toprule
Measure & Fricatives & Stops & Affricates & Nasals & Approximants \\
\midrule
Cells & 2{,}256 & 2{,}256 & 1{,}667 & 422 & 447 \\
\midrule
\auc, spectral & 0.95 & 0.83 & 0.85 & 0.71 & 0.76 \\
\auc, MFCC & 0.91 (0.90--0.91) & 0.87 (0.86--0.88) & 0.87 (0.85--0.88) & 0.88 (0.87--0.90) & 0.83 (0.82--0.85) \\
\auc, LFCC & 0.98 (0.98--0.99) & 0.95 (0.95--0.96) & 0.96 (0.96--0.97) & 0.95 (0.94--0.96) & 0.92 (0.91--0.93) \\
\midrule
\eer, MFCC & 0.17 & 0.20 & 0.20 & 0.18 & 0.24 \\
\eer, LFCC & 0.06 & 0.11 & 0.09 & 0.11 & 0.14 \\
\balacc, MFCC & 0.83 & 0.80 & 0.79 & 0.81 & 0.76 \\
\balacc, LFCC & 0.94 & 0.89 & 0.90 & 0.89 & 0.85 \\
\bottomrule
\end{tabular}}
\end{table}

\subsubsection{By context}

\paragraph{Obstruents}
Fricatives are classified best in all three contexts (Table~\ref{tab:p2-split}).
Among fricatives, the VC context has the highest \auc (0.96) and the CV context the lowest (0.92).
Fricatives in the isolated context lie between them (0.95).
For stops and affricates the CV context has the highest \auc (0.85 for both).
The gap between fricatives and the other classes is smallest in the CV context.
These orderings are descriptive, and we compute no intervals for them.

\paragraph{Sonorants}
Isolated tokens are detected least well for both classes (Table~\ref{tab:p3-context}).
The VC and CV values are 0.79 and 0.77 for approximants and 0.72 and 0.72 for nasals.
We compute no interval for these differences.

\subsubsection{By phoneme class and voicing}

\paragraph{Obstruents}
Voiceless cells have a higher \auc than voiced cells in all three manner classes (Table~\ref{tab:p2-split}).
The values are 0.95 and 0.94 for fricatives, 0.84 and 0.80 for stops, and 0.86 and 0.83 for affricates.
This agrees with Experiment 1, where voiceless tests were also significant more often in every manner class.
For affricates the two experiments are not directly comparable, because the Experiment 1 tests have little power (median 23 token pairs).

Table~\ref{tab:mv-results} puts the results of Experiments 1 and 3 side by side for each manner class and voicing category.
In all three manner classes, voiceless tokens have the higher share of significant tests and the higher classifier scores.

We also trained a classifier for each phoneme (Table~\ref{tab:phoneme}).
Among fricatives, /s/ is detected best (\auc 0.97), followed by /z/ (0.96), /f/ (0.95), /\textipa{S}/ (0.92), /v/ (0.92), /\textipa{T}/ (0.89), /\textipa{D}/ (0.89), /\textipa{Z}/ (0.88) and /h/ (0.86).
The /\textipa{Z}/ value rests on 6 cells.
Among stops, /t/ is detected best (0.85) and /g/ least well (0.77).
Every fricative phoneme scores higher than every stop phoneme.
The score of /h/ (0.86), the lowest among fricatives, is close to that of /t/, the best stop.
The fricative lead is therefore not carried by /s/ alone.

\paragraph{Sonorants}
All sonorants are voiced, so voicing does not vary.
XGBoost detects approximants better than nasals (0.76 against 0.71).
We also analysed the four approximant phonemes separately (Table~\ref{tab:p3-phoneme}).
Their mean \auc lies between 0.75 and 0.79, against 0.76 for the class as a whole.
We did not test the phoneme means against one another, and we do not interpret the order of phonemes that differ by a few hundredths.

\subsubsection{By system}

\paragraph{Obstruents}
For fricatives, conventional systems (\auc 0.95) and voice-cloning systems (0.95) are detected equally well (Table~\ref{tab:p2-system-group}).
For stops and affricates the cloning systems are detected less well than the conventional systems: 0.79 against 0.84 for stops, and 0.83 against 0.86 for affricates.
On the ten speakers that both groups share, the conventional systems score higher than the cloning systems in all three classes.
The values are 0.96 against 0.95 for fricatives, 0.84 against 0.79 for stops and 0.86 against 0.83 for affricates.
LJSpeech systems are harder to detect than VCTK systems in all three classes.

Individual systems span a wider range of \auc than the groups (Table~\ref{tab:p2-system}).
We do not compare the systems with a test, and the orderings are descriptive.
Among the VCTK systems, DiffGAN-TTS has the highest \auc in all three classes (0.97, 0.91 and 0.93).
VITS has the lowest \auc in all three classes (0.93 for fricatives, 0.77 for stops and 0.75 for affricates).
On LJSpeech, Transformer~TTS with HiFi-GAN is detected least well for fricatives (0.86) and affricates (0.74).
VITS and Transformer~TTS are detected least well for stops (0.62 for both).
Fricatives score above 0.85 for every system.
The LJSpeech values rest on 6 to 12 cells from a single speaker, and the Transformer~TTS value rests on 99 synthesised utterances.

\paragraph{Sonorants}
DiffGAN-TTS is detected best for both classes (Table~\ref{tab:p3-system}).
VITS is detected least well among the VCTK systems (0.62 for nasals and 0.65 for approximants).
LJSpeech VITS is also among the least detected systems: its \auc is 0.57 for nasals and 0.62 for approximants, against 0.62 and 0.70 for the eight LJSpeech systems together (Table~\ref{tab:p3-extra-auc}).
The LJSpeech VITS values rest on three cells each, and the pooled LJSpeech values on 24 cells each.
Cloning systems have a higher mean \auc than conventional systems for nasals and a slightly lower one for approximants.
On the ten shared speakers the same pattern holds (Table~\ref{tab:p3-extra-auc}).
We compute no interval for these group differences and do not test them.

\subsubsection{By feature}

\paragraph{Obstruents}
We measured feature contributions for fricatives.
Spectral shape contributes most, followed by spectral tilt, peak frequency, RMS amplitude, dynamic amplitude and duration (Table~\ref{tab:shap}).
Peak frequency ranks last by significance in Experiment 1 (Table~\ref{tab:p1-feature}) and third by SHAP value, so the two rankings differ for this feature.

\paragraph{Recording level}
Removing RMS amplitude lowers the mean \auc only slightly for fricatives (Table~\ref{tab:p2-norms}).
The mean change is $-0.008$, with an interval of $-0.012$ to $-0.006$.
Only 0.03 of the fricative cells lose more than 0.05 in \auc.
Average precision, equal error rate and balanced accuracy also change little.

The loss is larger for the other classes.
It is $-0.021$ for stops and $-0.031$ for affricates.
In 0.12 of the stop cells and 0.31 of the affricate cells the \auc falls by more than 0.05.
Stops and affricates therefore depend more on level than fricatives do.

The order of the classes does not change.
Without RMS the mean \auc is 0.94 for fricatives, 0.82 for affricates and 0.81 for stops.
Fricatives lose least, so their lead grows.
These gaps are differences of class means over all cells of each class, not over the shared cells of Table~\ref{tab:p2-matched}.
The gap in mean \auc between fricatives and affricates grows from 0.093 to 0.116, and between fricatives and stops from 0.120 to 0.133.
The largest drops are on LJSpeech systems with a HiFi-GAN vocoder, above all FastSpeech~2.
FastSpeech~2 with HiFi-GAN loses 0.04 for fricatives, 0.10 for stops and 0.14 for affricates.
These LJSpeech values rest on 6 to 12 cells per system.
For fricatives the mean loss is 0.02 on LJSpeech and 0.01 on VCTK.

\paragraph{Sonorants}
We measured feature contributions with SHAP values of the XGBoost classifiers on their test tokens, averaged over cells (Table~\ref{tab:p3-shap}).
For nasals the low-band energy ratio and the frequency of F1 contribute most, and duration and the antiformant features least.
For approximants the mean of F5 and the bandwidth of F3 contribute most, followed by the mean of F4.
Duration, the F2 slope and the F2--F3 distance contribute least.
Contributions are spread over several features.
No single feature is the largest contributor in more than 0.36 of the cells.
The two nasal features that are significant most often in the statistics, spectral tilt and the low-band energy ratio, rank third and first here.
The approximant feature that is significant most often, the bandwidth of F3, ranks second.

\subsubsection{Consistency and generalisation}

\paragraph{Obstruents}
The analyses in this paragraph cover fricatives.
The paragraph ``All classes'' compares all five classes.

Over the 30 VCTK speakers, Table~\ref{tab:p2-speaker} gives the minimum, median and maximum of the mean \auc per speaker, in the cells that pool all voicing categories and contexts.
The per-speaker means stay above 0.9.
We ran no test of speaker differences, so we make no claim about how much speakers differ.

Within VCTK, detection degrades as the test cell shares less with the training cell (Table~\ref{tab:p2-gen}).
The reference for each drop is the within-cell \auc, which is 0.96 for VCTK.
A change of system for the same speaker lowers the \auc by 0.03, to 0.93.
A change of speaker lowers it by 0.10, to 0.86.
A change of both lowers it by 0.13, to 0.83.
In the cross-system condition the human tokens, and therefore the sentences, are shared between the training and the test cell.
The cross-system \auc is therefore optimistic, and 0.03 is a lower bound on the effect of a change of system.
We report no matched within-cell reference for LJSpeech.
The mean over all 96 LJSpeech cells is 0.91 (Table~\ref{tab:p2-system-group}), and it is computed over different cells from the 0.86 of the cross-system row.

Synthetic-class recall is lower in the cross-speaker and cross-both conditions than in the cross-system condition (Table~\ref{tab:p2-recall}).
Human-class recall changes little between the cross-speaker and cross-both conditions.
In cross-system pairs, human recall is not a held-out measure, because the human tokens are shared between the training and test cell.
We report no within-cell recall, so the table does not show how much of the drop in \auc comes from synthetic-class recall.

In the cross-corpus rows of Table~\ref{tab:p2-gen}, the \auc is below 0.5 in both directions (0.40 to 0.43).
This is based on 31 test cells and 60 pairs for cross-speaker transfer, and 188 test cells and 2{,}820 pairs for cross-both transfer.
At a threshold of 0.5, synthetic recall is 0.40 to 0.47 and human recall is 0.40 to 0.51 across the four rows (Table~\ref{tab:p2-cross}).
For LJSpeech training cells and VCTK test cells, the table gives speaker-cluster bootstrap intervals.
The opposite direction has a single test speaker and no interval.

A single-feature check shows that the direction of two features differs between the corpora.
RMS amplitude lies near 0.5 on both (0.51 and 0.48; Table~\ref{tab:p2-single}).
So does duration (0.47 and 0.49).
Spectral shape ranks synthetic fricatives above human ones on LJSpeech (\auc 0.62 when used alone as the score) and below them on VCTK (0.41).
Dynamic amplitude does the same (0.54 and 0.35).
Spectral tilt has the same direction on both corpora (0.59 and 0.56).
The human tokens of the two corpora also differ in these features.
The \auc of LJSpeech against VCTK human tokens is 0.64 for RMS amplitude and 0.62 for both spectral tilt and peak frequency.
We did not test whether the below-chance cross-corpus values follow from these direction differences.

\paragraph{Sonorants}
We tested transfer between corpora for VITS, the one system with the same identifier on both corpora (Table~\ref{tab:p3-vits}).
Transfer across corpora is at chance in both directions.
The \auc is 0.51 and 0.52 for nasals, and 0.49 and 0.48 for approximants.
Transfer between two VCTK speakers is also close to chance, at 0.54 for nasals and 0.53 for approximants.
So the failure to transfer is not specific to crossing corpora.
The sonorant cues of one voice do not carry over to another voice, within a corpus or across corpora.
For comparison, fricative classifiers trained on one VCTK speaker and tested on another reach 0.86 (Table~\ref{tab:p2-gen}).
The features and phoneme classes differ, so this comparison is descriptive.
Transfer to unseen cells for all systems is compared across classes in the last paragraph of this section.

\paragraph{All classes}
We tested generalisation to unseen cells for all five classes (Table~\ref{tab:gen-all}).
Fricative cues carry over best to new speakers.
Within VCTK, the \auc on a new speaker is 0.86 for fricatives, against 0.77 for stops, 0.77 for affricates, 0.55 for nasals and 0.58 for approximants.
On a new speaker and a new system it is 0.83 for fricatives, against 0.71 for stops, 0.71 for affricates, 0.52 for nasals and 0.54 for approximants.
A new system for the same speaker is the easiest condition for every class.
Fricatives (0.93) and affricates (0.90) score highest there, and both sonorant classes score 0.75.
On LJSpeech, with one speaker and 8 test cells, the new-system \auc for fricatives is 0.86.
Across the two corpora no class transfers.
The obstruent classes score below chance (0.36 to 0.47), and the sonorant classes score close to chance (0.48 to 0.53).

\begin{table}[!htbp]
\centering
\caption{Experiment 3: generalisation to unseen cells, mean \auc over test cells, for each phoneme class. The VCTK rows use 180 test cells per condition and the LJSpeech row 8 test cells. The last two rows give the range over the two directions of transfer between the corpora. The fricative column is the fricative result of Table~\ref{tab:p2-gen}. No intervals are computed.}
\label{tab:gen-all}
\resizebox{\ifdim\width>\textwidth\textwidth\else\width\fi}{!}{%
\begin{tabular}{lccccc}
\toprule
Test cell & Fricatives & Stops & Affricates & Nasals & Approximants \\
\midrule
\multicolumn{6}{l}{\textit{Within VCTK}} \\
New system & 0.93 & 0.82 & 0.90 & 0.75 & 0.75 \\
New speaker & 0.86 & 0.77 & 0.77 & 0.55 & 0.58 \\
New speaker and system & 0.83 & 0.71 & 0.71 & 0.52 & 0.54 \\
\midrule
\multicolumn{6}{l}{\textit{Within LJSpeech}} \\
New system & 0.86 & 0.63 & 0.74 & 0.58 & 0.63 \\
\midrule
\multicolumn{6}{l}{\textit{Across corpora (range over the two directions)}} \\
New speaker & 0.40 & 0.41--0.46 & 0.36--0.41 & 0.51--0.53 & 0.48 \\
New speaker and system & 0.42--0.43 & 0.42--0.47 & 0.38 & 0.50--0.52 & 0.49--0.51 \\
\bottomrule
\end{tabular}}
\end{table}

\section{Discussion}
\label{sec:discussion}

\subsection{Fricatives are the place to look}

This study asked whether fricatives are the most separable class for telling synthetic from human speech.
For the systems and corpora we tested, the answer is yes.
Table~\ref{tab:evidence} brings together the evidence of the three experiments.

\begin{table}[!htbp]
\centering
\caption{Evidence by phoneme class. The significant share and the mean absolute Cliff's delta come from Experiments 1 and 2. For obstruents, they pool voiced and voiceless tests over all contexts and eight features. For sonorants, they use the CV and VC contexts and the sonorant formant features, so these two columns are not like for like. \auc is the Experiment 3 mean over all cells. ``Without RMS'' is the mean over the same cells with RMS amplitude removed, which applies to obstruents only. ``Cloning'' is the mean over the cells of the three voice-cloning systems. ``New speaker'' is the mean over test cells within VCTK for a classifier trained on another speaker of the same system. ``MFCC'' and ``LFCC'' are the class-neutral baselines. The last three rows count the system configurations, of 16, in which the class has the highest \auc, for the spectral features, MFCC and LFCC.}
\label{tab:evidence}
\begin{tabular}{lccccc}
\toprule
Measure & Fricatives & Stops & Affricates & Nasals & Approximants \\
\midrule
Significant tests & 0.69 & 0.65 & 0.16 & 0.40 & 0.43 \\
Mean $|\delta|$ & 0.22 & 0.18 & 0.35 & 0.12 & 0.09 \\
\midrule
\auc, spectral features & 0.95 & 0.83 & 0.85 & 0.71 & 0.76 \\
\auc without RMS & 0.94 & 0.81 & 0.82 & -- & -- \\
\auc, cloning systems & 0.95 & 0.79 & 0.83 & 0.78 & 0.76 \\
\auc, new speaker & 0.86 & 0.77 & 0.77 & 0.55 & 0.58 \\
\auc, MFCC & 0.91 & 0.87 & 0.87 & 0.88 & 0.83 \\
\auc, LFCC & 0.98 & 0.95 & 0.96 & 0.95 & 0.92 \\
\midrule
Top class in (of 16), spectral & 15 & 0 & 1 & 0 & 0 \\
Top class in (of 16), MFCC & 10 & 0 & 3 & 3 & 0 \\
Top class in (of 16), LFCC & 14 & 0 & 1 & 1 & 0 \\
\bottomrule
\end{tabular}
\end{table}

The classifiers single out fricatives.
The six obstruent features separate synthetic from human fricatives with a mean \auc of 0.95.
The same features give 0.85 for affricates and 0.83 for stops.
Sonorants, with their formant features and duration, reach 0.76 for approximants and 0.71 for nasals.
The paired comparison shows that the gap between the obstruent classes does not come from different sets of cells.
Fricatives are ahead of affricates by 0.09 and ahead of stops by 0.11.

With the class-specific features, the lead is broad.
It holds in every context and for both voicing categories.
It holds in 15 of the 16 system configurations.
The exception is FastSpeech~2 with HiFi-GAN on LJSpeech, where affricates score 0.96 and fricatives 0.94.
The affricate value rests on 11 cells with small test sets.

The lead depends on the representation.
With class-neutral features all classes are highly separable.
LFCC give 0.98 for fricatives and 0.92 to 0.96 for the other classes, and fricatives stay ahead of every class.
MFCC give 0.91 for fricatives and 0.83 to 0.88 for the others, and fricatives tie with nasals.
The fricative lead is therefore clearest with the class-specific features, and smaller with class-neutral features.
It is largest with the obstruent features, which were developed for fricatives.

The lead holds for the newest systems.
With the class-specific features, the three voice-cloning systems score 0.95 for fricatives, the same as the conventional systems (0.95).
Every other class scores lower for them: 0.83 for affricates, 0.79 for stops, 0.78 for nasals and 0.76 for approximants.
Cloning a voice from a reference recording therefore does not remove the fricative signal.

The lead does not depend on recording level.
Without RMS amplitude, the one level-dependent feature among the six, the mean \auc for fricatives falls only from 0.95 to 0.94.
Stops and affricates fall further.

The lead does not depend on which tokens are compared.
When human and synthetic tokens are matched by sentence, word and phoneme position, the mean \auc for fricatives stays at 0.95.

The lead is not carried by one sound.
Every fricative phoneme scores higher than every stop phoneme, and /s/ is highest at 0.97.

With the class-specific features, fricative cues also generalise best.
On a new speaker, fricative classifiers reach 0.86, against 0.77 for stops, 0.77 for affricates, 0.55 for nasals and 0.58 for approximants.
Sonorant formant cues barely carry over to a new voice.

The statistics alone would not single out fricatives.
Fricative and stop tests are significant more often than sonorant tests (0.69 and 0.65, against 0.40 and 0.43), although the features differ.
Fricatives and stops are close to each other.
Affricates have the largest effect sizes (0.35) and the smallest share of significant tests (0.16).
The small share of significant tests may reflect the few token pairs for affricates.
The classifiers make the difference between the classes clear.

\subsection{What the result does not show}

The cues are specific to a corpus.
Within VCTK, fricative classifiers still work on a new speaker (\auc 0.86) and on a new speaker and system (0.83).
Across the two corpora they fall below chance, at 0.40 to 0.43.
Two of the features, spectral shape and dynamic amplitude, point in opposite directions in the two corpora.
Fricatives are therefore the place to look, but a rule learned on one corpus is not a rule for another.
What to look for must be learned in the target domain.

The lead is a statement about handcrafted acoustic features.
PhonemeDF reports that handcrafted spectral features highlight fricatives and diphthongs, while self-supervised embeddings emphasise plosives \cite{nallaguntla2026phonemedf}.
We compared the classes with handcrafted spectral features and with MFCC and LFCC baselines.
We did not test self-supervised embeddings, so a different representation might rank stops higher.
We also did not compare vowels.
PhonemeDF finds diphthongs among the most discriminative categories, and Chhibber et al.\ rank periodic vowels lower \cite{chhibber2026phonetically}.
The two studies do not agree on vowels, and our study cannot settle this.

The statistics and the classifiers disagree about the voice-cloning systems.
On the ten shared speakers, cloning systems differ from human speech less often than conventional systems for obstruents (significant in 0.39 of tests against 0.58) and more often for sonorants (0.50 against 0.38).
The classifiers still score fricatives at 0.95 for the cloning systems.
We did not test why the two experiments differ.

We did not test why fricatives are the best place to look.
One possible reason is that fricatives are noise-like segments with no periodic structure to guide a model.
Another is that their spectral shape depends on fine articulatory detail.
The experiments here cannot tell these apart.

The sonorant formant feature sets were weaker for detection than class-neutral features.
Class-neutral features give 0.88 for nasals and 0.83 for approximants with MFCC, and 0.95 and 0.92 with LFCC.
The sonorant scores of Experiments 2 and 3 therefore describe those feature sets, and they do not set an upper limit on how detectable sonorants are.
Sonorant classifiers also barely transfer to a new speaker within VCTK (\auc 0.55 for nasals and 0.58 for approximants), so their cues are strongly specific to a voice.

Stops and affricates are less certain.
Both depend more on recording level than fricatives do.
The affricate test sets are small, with a median of 24 tokens.

\subsection{Implications}

For evaluation, fricatives are the first place to look.
A segment-level check of a TTS system can start with its fricatives, because they separate synthetic from human speech best.
The ranking of systems by fricative detectability is also informative.
On VCTK, the fricative \auc is highest for DiffGAN-TTS (0.97) and lowest for VITS (0.93).

For detection, fricative segments carry the strongest signal of the classes we tested.
A detector that uses them has to be trained for its target corpus, because the cues do not transfer.

\subsection{Limitations}
\label{sec:limitations}

\paragraph{Scope}
We compared consonant classes only.
Vowels, diphthongs and segment boundaries remain untested, and another class may separate synthetic from human speech better than fricatives.
We grouped /h/ with the fricatives, and liquids and glides as approximants, and we did not compare sibilant and non-sibilant fricatives.

\paragraph{Features}
The six obstruent features were developed for fricatives and used unchanged for stops and affricates.
Features designed for stops, such as burst and closure measures, might raise the stop scores.
The sonorant formant features scored well below class-neutral features, so the sonorant results describe those feature sets.
The class-neutral baselines were MFCC and LFCC only.
We did not test self-supervised embeddings, and we did not repeat the unseen-cell, recording-level or matched-token analyses with the baselines.
Only the class-neutral baseline compares all classes on equal footing, because the obstruent and sonorant features differ by design.

\paragraph{Signal and channel}
All audio was resampled to 16\,kHz, so the results describe the band below 8\,kHz, although fricative energy extends higher.
Recording conditions and vocoders differ between corpora and systems, and we ran no full control for them.
The only human-to-human comparison is the single-feature comparison of the two corpora.
We did not normalise loudness and tested only the removal of RMS amplitude.
Other effects of recording level on the remaining features were not tested.
The 40\,ms analysis window is a fixed choice.

\paragraph{Coverage}
The voice-cloning result rests on three systems, two of them adapted builds, and ten speakers.
A different release or hardware could behave differently.
The LJSpeech values rest on 6 to 12 cells per system, so the count of 15 of 16 configurations has limited weight.
GlowTTS, GradTTS and PFlow run on LJSpeech only, which has a single speaker.
Whether the synthesised VCTK sentences were part of any system's training data is not established.
Affricate test sets are small, with a median of 24 tokens.

\paragraph{Statistics}
Each classifier score comes from one split.
We treat tokens within an utterance as independent and do not model clustering by speaker and utterance.
The human tokens of a speaker are shared by all cells of that speaker, so the cells are not independent, and the unseen-cell means carry no intervals.
Nothing controls multiplicity across speaker--system pairs.
We did not check the symmetry assumption of the signed-rank test, and we compute Cliff's delta unpaired beside a paired test.
A cluster-aware re-analysis has not been done.

\paragraph{Reproducibility}
Decoding settings and random seeds of the TTS systems, the Transformer~TTS vocoder file on LJSpeech and software versions are not reported.
The x-vector and style encoders used with the ESPnet2 systems are not cited.

\subsection{Future work}

Four steps would extend this work.
The first is to normalise loudness and to extract the features again, so that all level effects are controlled.
The second is to test other features for sonorants.
The third is to repeat the experiments on more corpora with several speakers each, to see which cues transfer.
The fourth is to test whether listeners notice the fricative differences.

\section{Conclusion}
\label{sec:conclusion}

This paper examined whether fricatives are the most discriminative class for separating synthetic from human speech.
We compared five phoneme classes across 31 speakers, two corpora and 11 TTS systems, three of which clone a voice from a reference recording.
Experiments 1 and 2 characterised the acoustic differences of obstruents and sonorants with paired statistics.
Experiment 3 trained one classifier per speaker and system on both groups.

Fricatives were the most separable class.
Their mean \auc was 0.95, against 0.71 to 0.85 for the other classes.
The lead held in every context and voicing category and in 15 of 16 system configurations.
It held for the three voice-cloning systems, when RMS amplitude was removed, and when tokens were matched by sentence and phoneme position.
Fricative cues also carried over to new speakers better than those of any other class (\auc 0.86, against 0.55 to 0.77).

The lead depends on the representation.
It persisted under LFCC features (0.98, against 0.92 to 0.96).
Under MFCC features, fricatives remained ahead of stops, affricates and approximants but tied with nasals.
The result therefore concerns handcrafted spectral features and cepstral baselines, not self-supervised embeddings, which we did not test.

The acoustic cues did not transfer between corpora.
Obstruent classifiers trained on one corpus scored below chance on the other, and sonorant classifiers scored at chance.
Detectors and segment-level evaluations must therefore learn or recalibrate their cues for each target domain.

Fricatives are therefore the most separable consonant class under the class-specific and LFCC features, and the natural first target for analysis and detection.

\section*{Ethics statement}
All speech in this study comes from two public corpora: VCTK (version 0.92, released under the Creative Commons Attribution 4.0 licence) and LJSpeech (released into the public domain).
No new recordings were made, and no human participants took part.
The synthetic speech, including the voice-cloned speech of VCTK speakers, was generated only to analyse the acoustics of synthetic speech.

\clearpage
\appendix
\section{Supplementary tables}
\label{app:tables}
\setcounter{table}{0}
\renewcommand{\thetable}{A.\arabic{table}}
The tables below give the breakdowns behind the results in the main text.

\begin{table}[!htbp]
\centering
\caption{Tokens by manner class and voicing category, by corpus. The phonemes of each group are listed. /h/ is grouped with the voiceless fricatives.}
\label{tab:mv-data}
\resizebox{\ifdim\width>\textwidth\textwidth\else\width\fi}{!}{%
\begin{tabular}{lllrrrr}
\toprule
Manner class & Voicing & Phonemes & VCTK human & VCTK synthetic & LJSpeech human & LJSpeech synthetic \\
\midrule
Fricatives & Voiced & /v, \textipa{D}, z, \textipa{Z}/ & 42{,}905 & 184{,}656 & 13{,}537 & 30{,}337 \\
 & Voiceless & /f, \textipa{T}, s, \textipa{S}, h/ & 46{,}427 & 199{,}162 & 15{,}060 & 33{,}796 \\
\midrule
Stops & Voiced & /b, d, g/ & 24{,}006 & 145{,}217 & 11{,}180 & 25{,}214 \\
 & Voiceless & /p, t, k/ & 41{,}922 & 254{,}884 & 19{,}174 & 43{,}431 \\
\midrule
Affricates & Voiced & /\textipa{dZ}/ & 1{,}887 & 11{,}503 & 810 & 1{,}796 \\
 & Voiceless & /\textipa{tS}/ & 1{,}530 & 9{,}442 & 720 & 1{,}603 \\
\bottomrule
\end{tabular}}
\end{table}

\begin{table}[!htbp]
\centering
\caption{Synthetic fricative tokens and analysed utterances per system. Utterances are counted per speaker and summed over speakers.}
\label{tab:data-system}
\begin{tabular}{lrrrrr}
\toprule
 & \multicolumn{3}{c}{VCTK} & \multicolumn{2}{c}{LJSpeech} \\
System & Speakers & Utterances & Tokens & Utterances & Tokens \\
\midrule
DiffGAN-TTS     & 30 & 13{,}368 & 63{,}416 & 512   & 6{,}315 \\
FastSpeech~2    & 30 & 13{,}337 & 63{,}167 & 1{,}326 & 16{,}090 \\
Tacotron~2      & 30 & 13{,}331 & 63{,}222 & 1{,}326 & 16{,}093 \\
Transformer~TTS & 30 & 13{,}331 & 63{,}069 & 99    & 1{,}313 \\
VITS            & 30 & 13{,}368 & 63{,}404 & 498   & 6{,}113 \\
CosyVoice~2     & 10 & 4{,}712  & 22{,}527 & --    & -- \\
FishSpeech      & 10 & 4{,}711  & 22{,}513 & --    & -- \\
Qwen3-TTS       & 10 & 4{,}712  & 22{,}500 & --    & -- \\
GlowTTS         & -- & --       & --       & 498   & 6{,}121 \\
GradTTS         & -- & --       & --       & 486   & 5{,}980 \\
PFlow           & -- & --       & --       & 498   & 6{,}108 \\
\bottomrule
\end{tabular}
\end{table}

\begin{table}[!htbp]
\centering
\caption{Share of human tokens in each context, by corpus and phoneme group. Sonorants pool nasals and approximants. Over both corpora, voiced tokens are 0.48 of the human fricative tokens, 0.37 of the stop tokens and 0.55 of the affricate tokens. All sonorants are voiced.}
\label{tab:context-share}
\begin{tabular}{llccc}
\toprule
Corpus & Phoneme group & CV & VC & Isolated \\
\midrule
VCTK & Fricatives & 0.44 & 0.42 & 0.14 \\
 & Stops & 0.37 & 0.43 & 0.20 \\
 & Affricates & 0.47 & 0.43 & 0.09 \\
 & Sonorants & 0.42 & 0.57 & 0.01 \\
\midrule
LJSpeech & Fricatives & 0.41 & 0.44 & 0.14 \\
 & Stops & 0.34 & 0.41 & 0.25 \\
 & Affricates & 0.40 & 0.52 & 0.08 \\
 & Sonorants & 0.37 & 0.62 & 0.02 \\
\bottomrule
\end{tabular}
\end{table}

\begin{table}[!htbp]
\centering
\caption{Share of synthetic tokens (\%) that have a human counterpart under the pairing key, by phoneme class and corpus. Counts of synthetic tokens are those before the formant validity filter.}
\label{tab:pairing}
\begin{tabular}{lcc}
\toprule
Phoneme class & VCTK & LJSpeech \\
\midrule
Fricative & 99.4 & 99.2 \\
Stop & 98.8 & 99.6 \\
Affricate & 98.8 & 98.0 \\
Nasal & 99.2 & 98.8 \\
Approximant & 97.8 & 97.3 \\
\bottomrule
\end{tabular}
\end{table}

\begin{table}[!htbp]
\centering
\caption{Formant validity filter: tokens before the filter and the percentage retained, by phoneme class, corpus and label.}
\label{tab:filter}
\begin{tabular}{llrcrc}
\toprule
 &  & \multicolumn{2}{c}{Human} & \multicolumn{2}{c}{Synthetic} \\
Class & Corpus & Before & Retained (\%) & Before & Retained (\%) \\
\midrule
Nasal & VCTK & 34{,}918 & 81.8 & 212{,}125 & 81.9 \\
 & LJSpeech & 16{,}668 & 95.6 & 37{,}495 & 97.2 \\
Approximant & VCTK & 39{,}051 & 99.7 & 237{,}414 & 99.7 \\
 & LJSpeech & 16{,}616 & 99.7 & 37{,}826 & 99.8 \\
\bottomrule
\end{tabular}
\end{table}

\begin{table}[!htbp]
\centering
\caption{Minimum-size rules by analysis. Counts are tokens.}
\label{tab:minsize}
\begin{tabular}{lp{0.5\textwidth}}
\toprule
Analysis & Rule \\
\midrule
Experiment 3, obstruent classifier cell & at least 10 tokens in each class, and at least 5 of each class on each side of the split \\
Experiment 3, unseen-cell generalisation & at least 30 tokens in each class of a cell \\
Experiment 3, sonorant classifier cell & at least 30 tokens, with 20 for training and 10 for testing \\
Experiment 3, per-phoneme approximant cell & at least 30 tokens, as for other cells \\
Experiment 3, VITS transfer cell & at least 30 tokens in each class \\
Experiments 1 and 2, statistical test & at least two matched pairs, and not all differences zero \\
\bottomrule
\end{tabular}
\end{table}

\begin{table}[!htbp]
\centering
\caption{Experiment 1: share of significant tests and mean absolute Cliff's delta by context, over eight features and all obstruent classes.}
\label{tab:p1-context}
\begin{tabular}{lcc}
\toprule
Context & Significant & Mean $|\delta|$ \\
\midrule
CV       & 0.56 & 0.24 \\
VC       & 0.55 & 0.23 \\
Isolated & 0.41 & 0.27 \\
\bottomrule
\end{tabular}
\end{table}

\begin{table}[!htbp]
\centering
\caption{Experiment 1: conventional and voice-cloning systems on the ten speakers they share, over eight features. The difference is conventional minus cloning, with a 95\% speaker-cluster bootstrap interval.}
\label{tab:p1-shared}
\begin{tabular}{llcccc}
\toprule
Subset & Measure & Conventional & Cloning & Difference & 95\% interval \\
\midrule
All obstruents & Significant      & 0.58 & 0.39 & 0.20 & 0.15--0.24 \\
All obstruents & Mean $|\delta|$  & 0.28 & 0.19 & 0.09 & 0.07--0.11 \\
Fricatives     & Significant      & 0.75 & 0.55 & 0.21 & 0.15--0.26 \\
Fricatives     & Mean $|\delta|$  & 0.24 & 0.14 & 0.10 & 0.09--0.11 \\
\bottomrule
\end{tabular}
\end{table}

\begin{table}[!htbp]
\centering
\caption{Experiment 1: significant tests and mean absolute Cliff's delta by VCTK system, over eight features. The first two value columns cover all obstruent classes and the last two cover fricatives only.}
\label{tab:p1-persystem}
\begin{tabular}{lccccc}
\toprule
System & Speakers & Significant & Mean $|\delta|$ & Significant (fric.) & Mean $|\delta|$ (fric.) \\
\midrule
DiffGAN-TTS     & 30 & 0.56 & 0.32 & 0.74 & 0.29 \\
Tacotron~2      & 30 & 0.60 & 0.30 & 0.77 & 0.27 \\
FastSpeech~2    & 30 & 0.54 & 0.27 & 0.73 & 0.24 \\
Transformer~TTS & 30 & 0.54 & 0.24 & 0.74 & 0.21 \\
FishSpeech      & 10 & 0.40 & 0.20 & 0.57 & 0.15 \\
CosyVoice~2     & 10 & 0.36 & 0.18 & 0.50 & 0.14 \\
Qwen3-TTS       & 10 & 0.40 & 0.18 & 0.58 & 0.14 \\
VITS            & 30 & 0.43 & 0.18 & 0.62 & 0.16 \\
\bottomrule
\end{tabular}
\end{table}

\begin{table}[!htbp]
\centering
\caption{Experiment 1: significant tests and mean absolute Cliff's delta by feature, pooled over all obstruent classes, and direction by feature for fricatives only. Direction is the median of the signed delta among significant fricative tests. A negative value means lower in synthetic speech.}
\label{tab:p1-feature}
\begin{tabular}{lccc}
\toprule
Feature & Significant & Mean $|\delta|$ & Direction (fricatives) \\
\midrule
Spectral shape      & 0.64 & 0.30 & $-0.27$ \\
RMS amplitude       & 0.64 & 0.29 & $-0.16$ \\
Burst amplitude     & 0.60 & 0.29 & $-0.18$ \\
Dynamic amplitude   & 0.55 & 0.28 & $-0.33$ \\
Peak amplitude      & 0.54 & 0.27 & $-0.21$ \\
Spectral tilt       & 0.56 & 0.25 & $+0.22$ \\
Duration            & 0.33 & 0.13 & $-0.11$ \\
Peak frequency      & 0.21 & 0.15 & $-0.12$ \\
\bottomrule
\end{tabular}
\end{table}

\begin{table}[!htbp]
\centering
\caption{Experiment 1: direction of the effect among significant fricative tests. Shares are proportions of the significant fricative tests, given to two decimals.}
\label{tab:p1-direction}
\begin{tabular}{lc}
\toprule
Quantity & Value \\
\midrule
Spectral tilt higher in synthetic speech & 0.95 \\
Sign matches system--feature median, 5 features & 0.94 \\
Sign matches system--feature median, 8 features & 0.93 \\
\bottomrule
\end{tabular}
\end{table}

\begin{table}[!htbp]
\centering
\caption{Experiment 1: agreement between the paired analysis and the unpaired Dunn's-test analysis on the LJSpeech obstruent tests.}
\label{tab:p1-dunn}
\begin{tabular}{lc}
\toprule
Quantity & Value \\
\midrule
Tests & 1{,}144 \\
Agreement on significance & 0.88 \\
Correlation of Cliff's delta values & 1.00 \\
\bottomrule
\end{tabular}
\end{table}

\begin{table}[!htbp]
\centering
\caption{Experiment 2: median number of matched token pairs per test and share of significant tests, by context subset (nasals and approximants together unless stated).}
\label{tab:p3-pairs}
\begin{tabular}{lcc}
\toprule
Subset & Median matched pairs & Significant \\
\midrule
Isolated context & 10 & 0.005 \\
CV and VC contexts & 536 & 0.41 \\
Nasals, CV context & 185.5 & 0.30 \\
Nasals, VC context & 693.5 & 0.49 \\
\bottomrule
\end{tabular}
\end{table}

\begin{table}[!htbp]
\centering
\caption{Experiment 2: significant tests and mean absolute Cliff's delta by context.}
\label{tab:p3-stat-context}
\begin{tabular}{llccc}
\toprule
Class & Context & Tests & Significant & Mean $|\delta|$ \\
\midrule
Nasal & CV & 1{,}504 & 0.30 & 0.13 \\
Nasal & VC & 1{,}504 & 0.49 & 0.11 \\
Approximant & CV & 1{,}692 & 0.45 & 0.08 \\
Approximant & VC & 1{,}692 & 0.41 & 0.09 \\
\bottomrule
\end{tabular}
\end{table}

\begin{table}[!htbp]
\centering
\caption{Experiment 2: conventional and voice-cloning systems on the ten speakers they share, CV and VC contexts. The difference is conventional minus cloning, with a 95\% speaker-cluster bootstrap interval.}
\label{tab:p3-shared}
\begin{tabular}{llcccc}
\toprule
Class & Measure & Conventional & Cloning & Difference & 95\% interval \\
\midrule
All sonorants & Significant & 0.38 & 0.50 & $-$0.13 & $-$0.19 to $-$0.07 \\
All sonorants & Mean $|\delta|$ & 0.09 & 0.12 & $-$0.03 & $-$0.05 to $-$0.02 \\
Nasal & Significant & 0.38 & 0.51 & $-$0.13 & $-$0.18 to $-$0.07 \\
Nasal & Mean $|\delta|$ & 0.11 & 0.16 & $-$0.05 & $-$0.08 to $-$0.03 \\
Approximant & Significant & 0.38 & 0.50 & $-$0.13 & $-$0.21 to $-$0.05 \\
Approximant & Mean $|\delta|$ & 0.08 & 0.09 & $-$0.01 & $-$0.03 to 0.00 \\
\bottomrule
\end{tabular}
\end{table}

\begin{table}[!htbp]
\centering
\caption{Experiment 2: significant tests and mean absolute Cliff's delta by VCTK system, CV and VC contexts. Each nasal row covers 16 tests per speaker and each approximant row 18.}
\label{tab:p3-stat-persystem}
\begin{tabular}{lccccc}
\toprule
 & & \multicolumn{2}{c}{Nasal} & \multicolumn{2}{c}{Approximant} \\
System & Speakers & Significant & $|\delta|$ & Significant & $|\delta|$ \\
\midrule
DiffGAN-TTS & 30 & 0.64 & 0.22 & 0.64 & 0.15 \\
CosyVoice~2 & 10 & 0.58 & 0.19 & 0.54 & 0.10 \\
Qwen3-TTS & 10 & 0.48 & 0.15 & 0.44 & 0.07 \\
FishSpeech & 10 & 0.46 & 0.14 & 0.52 & 0.10 \\
FastSpeech~2 & 30 & 0.35 & 0.10 & 0.46 & 0.09 \\
Tacotron~2 & 30 & 0.34 & 0.10 & 0.37 & 0.08 \\
Transformer~TTS & 30 & 0.29 & 0.09 & 0.39 & 0.07 \\
VITS & 30 & 0.25 & 0.08 & 0.21 & 0.05 \\
\bottomrule
\end{tabular}
\end{table}

\begin{table}[!htbp]
\centering
\caption{Experiment 2: significant tests and mean absolute Cliff's delta by feature, CV and VC contexts.}
\label{tab:p3-stat-feature}
\begin{tabular}{lcccc}
\toprule
 & \multicolumn{2}{c}{Nasal} & \multicolumn{2}{c}{Approximant} \\
Feature & Significant & $|\delta|$ & Significant & $|\delta|$ \\
\midrule
\multicolumn{5}{l}{\emph{Nasal features}} \\
Spectral tilt (200--2000\,Hz) & 0.55 & 0.17 &  &  \\
Low-band energy ratio & 0.52 & 0.17 &  &  \\
Duration & 0.51 & 0.10 &  &  \\
F1 bandwidth & 0.41 & 0.13 &  &  \\
A1--P0 & 0.40 & 0.13 &  &  \\
F1 frequency & 0.37 & 0.13 &  &  \\
Antiformant frequency & 0.22 & 0.06 &  &  \\
Antiformant depth & 0.18 & 0.08 &  &  \\
\midrule
\multicolumn{5}{l}{\emph{Approximant features}} \\
F3 bandwidth &  &  & 0.65 & 0.18 \\
F3 mean &  &  & 0.52 & 0.08 \\
F4 mean &  &  & 0.51 & 0.10 \\
Duration &  &  & 0.48 & 0.07 \\
F2 mean &  &  & 0.48 & 0.07 \\
F5 mean &  &  & 0.48 & 0.12 \\
F2--F3 distance &  &  & 0.47 & 0.06 \\
F3 slope &  &  & 0.17 & 0.05 \\
F2 slope &  &  & 0.12 & 0.04 \\
\bottomrule
\end{tabular}
\end{table}

\begin{table}[!htbp]
\centering
\caption{Experiment 2: direction of the effect among significant tests, CV and VC contexts. The share is that of significant tests in which the stated direction holds.}
\label{tab:p3-direction}
\begin{tabular}{llc}
\toprule
Class & Feature and direction & Share of significant tests \\
\midrule
Approximant & F3 bandwidth higher in synthetic speech & 0.85 \\
Nasal & F1 bandwidth higher in synthetic speech & 0.72 \\
Nasal & Low-band energy ratio higher in synthetic speech & 0.73 \\
Nasal & Antiformant depth lower in synthetic speech & 0.65 \\
Nasal & Antiformant frequency lower in synthetic speech & 0.66 \\
Approximant & F2 mean higher in synthetic speech & 0.47 \\
\bottomrule
\end{tabular}
\end{table}

\begin{table}[!htbp]
\centering
\caption{Experiment 3: the main analysis and the matched-token analysis, over the cells that both share. Each metric is given for the main analysis and then for the matched tokens. $\Delta$\auc is the mean change in \auc per cell with a 95\% speaker-cluster bootstrap interval. The last column is the median number of test tokens in a cell. Sonorant rows use XGBoost. Matching removes some sonorant cells, which fall under the size rule, so the rows compare the cells that remain.}
\label{tab:matched-tokens}
\resizebox{\ifdim\width>\textwidth\textwidth\else\width\fi}{!}{%
\begin{tabular}{lccccccc}
\toprule
Manner class & Cells & \auc & \ap & \eer & \balacc & $\Delta$\auc & Test tokens \\
\midrule
Fricatives & 2{,}256 & 0.95 / 0.95 & 0.93 / 0.93 & 0.12 / 0.12 & 0.88 / 0.88 & $-$0.0004 ($-$0.0013 to 0.0007) & 379 / 354 \\
Stops & 2{,}256 & 0.83 / 0.83 & 0.81 / 0.82 & 0.25 / 0.25 & 0.75 / 0.75 & $-$0.0012 ($-$0.0026 to 0.0002) & 312.5 / 304 \\
Affricates & 1{,}644 & 0.85 / 0.85 & 0.85 / 0.86 & 0.21 / 0.21 & 0.78 / 0.78 & 0.0000 ($-$0.0034 to 0.0034) & 25 / 24 \\
Nasals & 398 & 0.71 / 0.71 & 0.69 / 0.71 & 0.34 / 0.34 & 0.66 / 0.66 & $-$0.0064 ($-$0.0117 to $-$0.0013) & 133 / 123 \\
Approximants & 387 & 0.78 / 0.78 & 0.74 / 0.76 & 0.29 / 0.29 & 0.71 / 0.71 & 0.0004 ($-$0.0021 to 0.0031) & 303 / 290 \\
\bottomrule
\end{tabular}}
\end{table}

\begin{table}[!htbp]
\centering
\caption{Experiment 3: paired comparison of fricatives with each other class under the class-neutral baseline. The gap is the mean difference in \auc (fricatives minus the other class) over the cells the two classes share, with a 95\% speaker-cluster bootstrap interval. The last column is the share of shared cells in which fricatives have the higher \auc. Sonorant cells are paired on speaker, system, the voiced category and context.}
\label{tab:cn-gaps}
\resizebox{\ifdim\width>\textwidth\textwidth\else\width\fi}{!}{%
\begin{tabular}{llcccc}
\toprule
Representation & Compared with & Cells & Mean gap & 95\% interval & Share of cells \\
\midrule
MFCC & Stops & 2{,}256 & 0.036 & 0.031 to 0.043 & 0.80 \\
 & Affricates & 1{,}667 & 0.040 & 0.030 to 0.049 & 0.61 \\
 & Nasals & 422 & 0.004 & $-$0.011 to 0.022 & 0.41 \\
 & Approximants & 447 & 0.053 & 0.041 to 0.064 & 0.70 \\
\midrule
LFCC & Stops & 2{,}256 & 0.028 & 0.024 to 0.032 & 0.90 \\
 & Affricates & 1{,}667 & 0.020 & 0.015 to 0.025 & 0.54 \\
 & Nasals & 422 & 0.024 & 0.016 to 0.032 & 0.69 \\
 & Approximants & 447 & 0.055 & 0.046 to 0.064 & 0.85 \\
\bottomrule
\end{tabular}}
\end{table}

\begin{table}[!htbp]
\centering
\caption{Experiment 3: XGBoost \auc by context.}
\label{tab:p3-context}
\begin{tabular}{lcc}
\toprule
Context & Nasal & Approximant \\
\midrule
CV & 0.72 & 0.77 \\
VC & 0.72 & 0.79 \\
Isolated & 0.64 & 0.66 \\
\bottomrule
\end{tabular}
\end{table}

\begin{table}[!htbp]
\centering
\caption{Experiment 3: mean \auc over all cells carrying each context label or each voicing label. The context labels and the voicing labels cover different sets of cells: for fricatives and stops each context label covers 564 cells and each voicing label 752 cells. For affricates the counts are smaller, and the isolated context has only 26 cells. Cells that pool all contexts or both voicing categories are not tabulated separately, so the headline means cannot be rebuilt from this table.}
\label{tab:p2-split}
\begin{tabular}{lccccc}
\toprule
 & \multicolumn{3}{c}{Context} & \multicolumn{2}{c}{Voicing} \\
Manner class & CV & VC & Isolated & Voiced & Voiceless \\
\midrule
Fricatives & 0.92 & 0.96 & 0.95 & 0.94 & 0.95 \\
Stops & 0.85 & 0.81 & 0.81 & 0.80 & 0.84 \\
Affricates & 0.85 & 0.84 & 0.81 & 0.83 & 0.86 \\
\bottomrule
\end{tabular}
\end{table}

\begin{table}[!htbp]
\centering
\caption{Obstruent results by manner class and voicing category. Experiment 1 columns give the number of tests, the share that are significant and the mean absolute Cliff's delta. Experiment 3 columns give the number of cells carrying the voicing label and the mean classifier scores over those cells.}
\label{tab:mv-results}
\resizebox{\ifdim\width>\textwidth\textwidth\else\width\fi}{!}{%
\begin{tabular}{llcccccccc}
\toprule
 & & \multicolumn{3}{c}{Experiment 1} & \multicolumn{5}{c}{Experiment 3} \\
Manner class & Voicing & Tests & Significant & Mean $|\delta|$ & Cells & \auc & \ap & \eer & \balacc \\
\midrule
Fricatives & Voiced & 4{,}512 & 0.63 & 0.23 & 752 & 0.94 & 0.92 & 0.13 & 0.87 \\
 & Voiceless & 4{,}512 & 0.75 & 0.21 & 752 & 0.95 & 0.93 & 0.11 & 0.88 \\
\midrule
Stops & Voiced & 4{,}512 & 0.58 & 0.18 & 752 & 0.80 & 0.79 & 0.27 & 0.73 \\
 & Voiceless & 4{,}512 & 0.72 & 0.18 & 752 & 0.84 & 0.82 & 0.24 & 0.76 \\
\midrule
Affricates & Voiced & 4{,}504 & 0.15 & 0.33 & 548 & 0.83 & 0.83 & 0.23 & 0.75 \\
 & Voiceless & 3{,}973 & 0.17 & 0.37 & 535 & 0.86 & 0.86 & 0.21 & 0.78 \\
\bottomrule
\end{tabular}}
\end{table}

\begin{table}[!htbp]
\centering
\caption{Experiment 3: classifier scores for each fricative and stop phoneme analysed separately, ordered by \auc within each class. A cell is one speaker, system and context. The /\textipa{Z}/ value rests on 6 cells.}
\label{tab:phoneme}
\begin{tabular}{llccccc}
\toprule
Manner class & Phoneme & Cells & \auc & \ap & \eer & \balacc \\
\midrule
Fricatives & /s/ & 752 & 0.97 & 0.97 & 0.07 & 0.92 \\
 & /z/ & 570 & 0.96 & 0.95 & 0.09 & 0.90 \\
 & /f/ & 682 & 0.95 & 0.93 & 0.10 & 0.89 \\
 & /\textipa{S}/ & 564 & 0.92 & 0.90 & 0.14 & 0.85 \\
 & /v/ & 564 & 0.92 & 0.89 & 0.15 & 0.85 \\
 & /\textipa{T}/ & 577 & 0.89 & 0.87 & 0.17 & 0.81 \\
 & /\textipa{D}/ & 515 & 0.89 & 0.86 & 0.18 & 0.81 \\
 & /\textipa{Z}/ & 6 & 0.88 & 0.82 & 0.20 & 0.82 \\
 & /h/ & 430 & 0.86 & 0.81 & 0.21 & 0.78 \\
\midrule
Stops & /t/ & 752 & 0.85 & 0.84 & 0.22 & 0.78 \\
 & /p/ & 752 & 0.81 & 0.80 & 0.26 & 0.74 \\
 & /k/ & 752 & 0.81 & 0.80 & 0.26 & 0.74 \\
 & /d/ & 752 & 0.81 & 0.79 & 0.26 & 0.74 \\
 & /b/ & 687 & 0.79 & 0.79 & 0.27 & 0.72 \\
 & /g/ & 685 & 0.77 & 0.77 & 0.29 & 0.70 \\
\bottomrule
\end{tabular}
\end{table}

\begin{table}[!htbp]
\centering
\caption{Experiment 3: XGBoost \auc for approximant phonemes analysed separately, with the number of cells that have a defined \auc.}
\label{tab:p3-phoneme}
\begin{tabular}{lcc}
\toprule
Phoneme & Cells & \auc \\
\midrule
/l/ & 376 & 0.79 \\
/\textipa{r}/ & 387 & 0.78 \\
/j/ & 188 & 0.76 \\
/w/ & 203 & 0.75 \\
\bottomrule
\end{tabular}
\end{table}

\begin{table}[!htbp]
\centering
\caption{Experiment 3: mean \auc by system group and manner class, with the number of cells in brackets. The last row restricts the conventional systems to the ten speakers covered by the cloning systems.}
\label{tab:p2-system-group}
\begin{tabular}{lccc}
\toprule
System group & Fricatives & Stops & Affricates \\
\midrule
Conventional VCTK (30 speakers) & 0.95 (1{,}800) & 0.84 (1{,}800) & 0.86 (1{,}313) \\
Voice cloning (10 speakers) & 0.95 (360) & 0.79 (360) & 0.83 (272) \\
LJSpeech & 0.91 (96) & 0.72 (96) & 0.84 (82) \\
\midrule
Conventional VCTK (same 10 speakers) & 0.96 (600) & 0.84 (600) & 0.86 (457) \\
\bottomrule
\end{tabular}
\end{table}

\begin{table}[!htbp]
\centering
\caption{Experiment 3: mean \auc by TTS system and manner class. $n$ is the number of cells for fricatives and stops, which are equal, and for affricates. No intervals are computed for the per-system values.}
\label{tab:p2-system}
\resizebox{\ifdim\width>\textwidth\textwidth\else\width\fi}{!}{%
\begin{tabular}{lccccc}
\toprule
System & $n$ (fric., stop) & Fricative & Stop & $n$ (aff.) & Affricate \\
\midrule
DiffGAN-TTS (VCTK) & 360 & 0.97 & 0.91 & 261 & 0.93 \\
FishSpeech (VCTK) & 120 & 0.96 & 0.81 & 91 & 0.86 \\
Tacotron~2 (VCTK) & 360 & 0.96 & 0.87 & 264 & 0.91 \\
Qwen3-TTS (VCTK) & 120 & 0.95 & 0.78 & 91 & 0.81 \\
Transformer~TTS (VCTK) & 360 & 0.95 & 0.83 & 262 & 0.85 \\
CosyVoice~2 (VCTK) & 120 & 0.94 & 0.78 & 90 & 0.81 \\
FastSpeech~2 (VCTK) & 360 & 0.94 & 0.84 & 262 & 0.87 \\
VITS (VCTK) & 360 & 0.93 & 0.77 & 264 & 0.75 \\
\midrule
Tacotron~2 + HiFi-GAN (LJS) & 12 & 0.94 & 0.79 & 11 & 0.86 \\
FastSpeech~2 + HiFi-GAN (LJS) & 12 & 0.94 & 0.83 & 11 & 0.96 \\
DiffGAN-TTS + HiFi-GAN (LJS) & 12 & 0.94 & 0.81 & 10 & 0.93 \\
PFlow + HiFi-GAN (LJS) & 12 & 0.90 & 0.72 & 11 & 0.81 \\
GradTTS + HiFi-GAN (LJS) & 12 & 0.90 & 0.68 & 11 & 0.80 \\
GlowTTS + HiFi-GAN (LJS) & 12 & 0.89 & 0.67 & 11 & 0.78 \\
VITS (LJS) & 12 & 0.89 & 0.62 & 11 & 0.76 \\
Transformer~TTS + HiFi-GAN (LJS) & 12 & 0.86 & 0.62 & 6 & 0.74 \\
\bottomrule
\end{tabular}}
\end{table}

\begin{table}[!htbp]
\centering
\caption{Experiment 3: XGBoost \auc by VCTK system, with the number of cells. No intervals are computed for the per-system values.}
\label{tab:p3-system}
\begin{tabular}{lcccc}
\toprule
System & Nasal cells & Nasal & Approximant cells & Approximant \\
\midrule
DiffGAN-TTS & 67 & 0.86 & 65 & 0.90 \\
CosyVoice~2 & 24 & 0.82 & 24 & 0.79 \\
FishSpeech & 22 & 0.75 & 23 & 0.78 \\
Qwen3-TTS & 22 & 0.75 & 21 & 0.71 \\
FastSpeech~2 & 66 & 0.68 & 72 & 0.78 \\
Transformer~TTS & 64 & 0.68 & 71 & 0.75 \\
Tacotron~2 & 67 & 0.68 & 73 & 0.76 \\
VITS & 66 & 0.62 & 74 & 0.65 \\
\bottomrule
\end{tabular}
\end{table}

\begin{table}[!htbp]
\centering
\caption{Experiment 3: XGBoost \auc for subsets quoted in the text. ``Shared'' restricts both groups to the ten speakers covered by the cloning systems. The pooled LJSpeech row covers eight systems with 24 cells per class, and the VITS row 3 cells per class.}
\label{tab:p3-extra-auc}
\begin{tabular}{llcc}
\toprule
Subset & Group & Nasal & Approximant \\
\midrule
Shared speakers & Voice cloning & 0.78 & 0.76 \\
Shared speakers & Conventional & 0.70 & 0.78 \\
LJSpeech & Eight systems pooled & 0.62 & 0.70 \\
LJSpeech & VITS only & 0.57 & 0.62 \\
\bottomrule
\end{tabular}
\end{table}

\begin{table}[!htbp]
\centering
\caption{Experiment 3: feature attribution for fricatives, as the absolute value of the per-cell signed mean SHAP value (log-odds units), averaged over cells. This is not the mean absolute SHAP value of Table~\ref{tab:p3-shap}.}
\label{tab:shap}
\begin{tabular}{lc}
\toprule
Feature & SHAP \\
\midrule
Spectral shape & 0.61 \\
Spectral tilt & 0.52 \\
Peak frequency & 0.26 \\
RMS amplitude & 0.24 \\
Dynamic amplitude & 0.11 \\
Duration & 0.08 \\
\bottomrule
\end{tabular}
\end{table}

\begin{table}[!htbp]
\centering
\caption{Experiment 3: the effect of removing RMS amplitude. Each metric is given with all six features and then with the five gain-invariant features, as a mean over the same cells. $\Delta$\auc is the mean change in \auc per cell with a 95\% speaker-cluster bootstrap interval. The last column is the share of cells whose \auc falls by more than 0.05.}
\label{tab:p2-norms}
\resizebox{\ifdim\width>\textwidth\textwidth\else\width\fi}{!}{%
\begin{tabular}{lccccccc}
\toprule
Manner class & Cells & \auc & \ap & \eer & \balacc & $\Delta$\auc & Share, fall $>0.05$ \\
\midrule
Fricatives & 2{,}256 & 0.95 / 0.94 & 0.93 / 0.92 & 0.12 / 0.13 & 0.88 / 0.87 & $-$0.008 ($-$0.012 to $-$0.006) & 0.03 \\
Stops & 2{,}256 & 0.83 / 0.81 & 0.81 / 0.79 & 0.25 / 0.27 & 0.75 / 0.73 & $-$0.021 ($-$0.026 to $-$0.018) & 0.12 \\
Affricates & 1{,}667 & 0.85 / 0.82 & 0.85 / 0.83 & 0.21 / 0.25 & 0.78 / 0.74 & $-$0.031 ($-$0.037 to $-$0.025) & 0.31 \\
\bottomrule
\end{tabular}}
\end{table}

\begin{table}[!htbp]
\centering
\caption{Experiment 3: mean absolute SHAP value (log-odds units) and share of cells in which the feature is the largest contributor, XGBoost classifiers. This is not the per-cell signed mean of Table~\ref{tab:shap}.}
\label{tab:p3-shap}
\begin{tabular}{lcc}
\toprule
Feature & Mean $|$SHAP$|$ & Cells where largest \\
\midrule
\multicolumn{3}{l}{\emph{Nasals (422 cells)}} \\
Low-band energy ratio & 0.83 & 0.27 \\
F1 frequency & 0.78 & 0.24 \\
Spectral tilt (200--2000\,Hz) & 0.69 & 0.17 \\
A1--P0 & 0.64 & 0.12 \\
F1 bandwidth & 0.62 & 0.11 \\
Antiformant frequency & 0.49 & 0.03 \\
Antiformant depth & 0.49 & 0.04 \\
Duration & 0.39 & 0.01 \\
\midrule
\multicolumn{3}{l}{\emph{Approximants (447 cells)}} \\
F5 mean & 0.79 & 0.36 \\
F3 bandwidth & 0.77 & 0.32 \\
F4 mean & 0.68 & 0.17 \\
F3 mean & 0.55 & 0.05 \\
F2 mean & 0.47 & 0.04 \\
F3 slope & 0.44 & 0.02 \\
F2--F3 distance & 0.38 & 0.01 \\
F2 slope & 0.38 & 0.02 \\
Duration & 0.26 & 0.01 \\
\bottomrule
\end{tabular}
\end{table}

\begin{table}[!htbp]
\centering
\caption{Experiment 3: mean \auc per speaker over the 30 VCTK speakers for fricatives, in cells that pool all voicing categories and contexts.}
\label{tab:p2-speaker}
\begin{tabular}{ccc}
\toprule
Minimum & Median & Maximum \\
\midrule
0.92 & 0.96 & 0.98 \\
\bottomrule
\end{tabular}
\end{table}

\begin{table}[!htbp]
\centering
\caption{Experiment 3: generalisation of fricative classifiers to unseen cells, averaged over test cells. The within-cell row is the reference from held-out tokens of the same cell. The rows that cross corpora give a range of \auc, because the value depends on the direction of transfer (VCTK to LJSpeech or the reverse).}
\label{tab:p2-gen}
\begin{tabular}{llccc}
\toprule
Corpus & Condition & Test cells & Pairs & \auc \\
\midrule
VCTK & Within-cell & -- & -- & 0.96 \\
VCTK & Cross-system & 180 & 960 & 0.93 \\
VCTK & Cross-speaker & 180 & 4{,}620 & 0.86 \\
VCTK & Cross-both & 180 & 26{,}640 & 0.83 \\
LJSpeech & Cross-system & 8 & 56 & 0.86 \\
\midrule
VCTK $\leftrightarrow$ LJSpeech & Cross-speaker & 31 & 60 & 0.40 \\
VCTK $\leftrightarrow$ LJSpeech & Cross-both & 188 & 2{,}820 & 0.42--0.43 \\
\bottomrule
\end{tabular}
\end{table}

\begin{table}[!htbp]
\centering
\caption{Experiment 3: recall of fricative classifiers at a threshold of 0.5 within VCTK, by condition. Human recall in the cross-system row is not a held-out measure.}
\label{tab:p2-recall}
\begin{tabular}{lcc}
\toprule
Condition & Synthetic recall & Human recall \\
\midrule
Cross-system & 0.73 & 0.96 \\
Cross-speaker & 0.68 & 0.85 \\
Cross-both & 0.62 & 0.85 \\
\bottomrule
\end{tabular}
\end{table}

\begin{table}[!htbp]
\centering
\caption{Experiment 3: cross-corpus transfer of fricative classifiers, by condition and direction. Recall is at a threshold of 0.5 and is the mean over pairs. The interval is a 95\% speaker-cluster bootstrap interval. A dash marks the direction with a single test speaker, for which no interval exists.}
\label{tab:p2-cross}
\begin{tabular}{llcccc}
\toprule
Condition & Training $\to$ test & \auc & 95\% interval & Synthetic recall & Human recall \\
\midrule
Cross-speaker & LJSpeech $\to$ VCTK & 0.40 & 0.38--0.42 & 0.47 & 0.40 \\
Cross-speaker & VCTK $\to$ LJSpeech & 0.40 & -- & 0.41 & 0.46 \\
Cross-both & LJSpeech $\to$ VCTK & 0.43 & 0.42--0.45 & 0.40 & 0.51 \\
Cross-both & VCTK $\to$ LJSpeech & 0.42 & -- & 0.43 & 0.49 \\
\bottomrule
\end{tabular}
\end{table}

\begin{table}[!htbp]
\centering
\caption{Experiment 3: \auc of each single feature used alone as the score, pooled over fricative tokens. The first two columns rank synthetic tokens against human tokens within one corpus. The last column ranks LJSpeech human tokens against VCTK human tokens. The tokens are the fricative tokens counted in Table~\ref{tab:data}.}
\label{tab:p2-single}
\resizebox{\ifdim\width>\textwidth\textwidth\else\width\fi}{!}{%
\begin{tabular}{lccc}
\toprule
Feature & LJSpeech, synthetic vs human & VCTK, synthetic vs human & LJSpeech vs VCTK, human \\
\midrule
RMS amplitude & 0.51 & 0.48 & 0.64 \\
Spectral tilt & 0.59 & 0.56 & 0.62 \\
Spectral shape & 0.62 & 0.41 & 0.43 \\
Dynamic amplitude & 0.54 & 0.35 & 0.40 \\
Peak frequency & 0.44 & 0.48 & 0.62 \\
Duration & 0.47 & 0.49 & 0.58 \\
\bottomrule
\end{tabular}}
\end{table}

\begin{table}[!htbp]
\centering
\caption{Experiment 3: transfer of VITS sonorant classifiers, mean \auc over pairs of cells in the CV and VC contexts. The corpus-crossing columns have 60 pairs each, and the VCTK to VCTK column has 1{,}740 pairs.}
\label{tab:p3-vits}
\resizebox{\ifdim\width>\textwidth\textwidth\else\width\fi}{!}{%
\begin{tabular}{lccc}
\toprule
Class & VCTK $\to$ LJSpeech & LJSpeech $\to$ VCTK & VCTK $\to$ VCTK, other speaker \\
\midrule
Nasal & 0.51 & 0.52 & 0.54 \\
Approximant & 0.49 & 0.48 & 0.53 \\
\bottomrule
\end{tabular}}
\end{table}

\clearpage
\section*{Data and code availability}
All code and data used in this study will be made publicly available upon publication.

\section*{Declaration of competing interest}
The authors declare that they have no known competing financial interests or personal relationships that could have appeared to influence the work reported in this paper.

\section*{CRediT authorship contribution statement}
\textbf{Sriyugesh Bhyravajjula:} Conceptualization, Methodology, Software, Validation, Formal analysis, Investigation, Resources, Data curation, Writing -- original draft, Writing -- review \& editing, Visualization, Funding acquisition.
\textbf{Ayushi Pandey:} Supervision, Project administration.
\textbf{Kevin Tang:} Supervision, Project administration.

\section*{Declaration of generative AI and AI-assisted technologies in the manuscript preparation process}
During the preparation of this work the authors used Claude (Anthropic) to draft and revise text from the authors' analyses, results and outline, to improve language and structure, and to check the manuscript against the journal's guidelines and verify its references.
After using this tool, the authors reviewed and edited the content as needed and take full responsibility for the content of the publication.

\section*{Funding}
This work was self-funded by the first author.
It did not receive any specific grant from funding agencies in the public, commercial, or not-for-profit sectors.

\end{document}